\documentclass[useAMS,usenatbib,letter]{mn2e}
\usepackage{times}
\usepackage{graphicx}
\usepackage{amssymb}
\usepackage{natbib}
\usepackage{color,ulem}
\bibpunct{(}{)}{;}{a}{}{,}

\title[Galaxy populations in Antlia. II.]{Galaxy populations
in the Antlia cluster. II. Compact elliptical galaxy candidates  
\thanks{Based on observations carried out at the Cerro Tololo Inter--American 
  Observatory (Chile), at Las Campanas Observatory (Chile), and 
  at the European Southern Observatory, Paranal (Chile) (Program 71.B-0122(A)). 
  Also based on observations made with the NASA/ESA Hubble Space Telescope, 
  obtained from the data archive at the Space Telescope Institute. STScI is 
  operated by the association of Universities for Research in Astronomy, Inc. 
  under the NASA contract NAS 5-26555.}}
\author[Smith Castelli et al.]{Anal\'ia V. Smith Castelli$^{1,2,3}$ 
\thanks{E-mail:\,asmith@fcaglp.unlp.edu.ar\,\,(ASC);\,favio@fcaglp.unlp.edu.ar\,(FF);
\,tom@mobydick.cfm.udec.cl\,\,(TR);\,lbassino@fcaglp.unlp.edu.ar\,(LB)} 
Favio R. Faifer$^{1,2,3}$, Tom Richtler$^{4}$ and Lilia P. Bassino$^{1,2,3}$\\  
$^{1}$Facultad de Ciencias Astron\'omicas y Geof\'{\i}sicas,
      Universidad Nacional de La Plata,
      Paseo del Bosque, B1900FWA La Plata,
      Argentina\\ 
$^{2}$Instituto de Astrof\'isica de La Plata (CCT La Plata - CONICET - UNLP) \\
$^{3}$Consejo Nacional de Investigaciones Cient\'ificas y T\'ecnicas, Rivadavia 1917, Buenos Aires, Argentina\\
$^{4}$Departamento de Astronom\'ia, Universidad de Concepci\'on, Casilla 160-C, 
Concepci\'on, Chile} 

\begin{document}

\date{Accepted . Received ; in original form }
\maketitle

\label{firstpage}

\begin{abstract}
Continuing our study of galaxy populations in the Antlia cluster,
we present a photometric analysis of four galaxies classified as 
compact elliptical (cE) galaxies in the Ferguson \& Sandage 
(1990, hereafter FS90) Antlia Group catalogue. Only 6 members of 
this rare type of galaxies are known until now.
Using data in various photometric systems (Washington $C$, Kron-Cousins $R$,
Bessel $V$ and $I$, HST $F814W$ and $F435W$),
we measured brightness and colour profiles, as well
as structural parameters. 
By comparing them with those of other galaxies 
in the Antlia cluster, as well as with confirmed cE galaxies  
from the literature, we  found that two of the FS90 cE candidates, 
albeit being spectroscopically confirmed Antlia members, are not cE galaxies.
However, one of these objects presents strong ellipticity and position 
angle variations that resemble those already reported for M32, leading us to
speculate about this kind of objects being progenitors of cE 
galaxies. The other two FS90 cE candidates, for which radial velocities are not 
available, match some features typical of cE galaxies like being close in 
projection to a larger galaxy, displaying flat colour profiles, and having a 
high degree of compactness. 
Only one of the remaining cE candidates shows a high central
surface brightness, two components in 
its brightness profile, distinct changes in ellipticity and 
position angle where the outer component begins to dominate,
and seems to follow the same trend as other confirmed 
cE galaxies in a luminosity versus mean effective surface brightness 
diagram. Moreover, it shows a distorted inner structure with similar 
characteristics to those found by simulations of interacting galaxies,
and an extremely faint structure that seems to link this object 
with one of the Antlia dominant galaxies, has been detected in MOSAIC-CTIO, 
FORS1-VLT, and ACS-HST images. The cE nature of this galaxy as 
well as the possible interaction with its bright companion, still have 
to be confirmed through spectroscopy. 
\end{abstract}

\begin{keywords}
galaxies: clusters: general -- galaxies: clusters: individual: Antlia -- 
galaxies: elliptical and lenticular -- galaxies: dwarf -- 
galaxies: photometry -- galaxies: peculiar 
\end{keywords}
\section{Introduction}
\label{intro}

Early-type dwarf galaxies are the most frequent galaxy type in nearby 
groups and clusters of galaxies \citep[e.g.][]{FS90}. Although the information 
related with their 
evolution in such environments is codified in their spatial distribution,
chemical composition and kinematics, a single scenario including all 
their properties in a consistent frame is still missing. The main 
difficulty resides in the fact that they do not constitute a 
homogeneous class of objects. Among them, there are nucleated and
non-nucleated galaxies, as well as systems that were found to harbour
discs and spiral structure \citep*{J00,CB05,L06b}. Moreover, 
central star formation has also been reported in some objects \citep{L06a}.

The so called compact elliptical \citep[cE, e.g.][]{N87}, compact dwarf 
elliptical \citep[cdE,][]{D01} or M32-like \citep[e.g.][]{ZB98} galaxies are 
members of the low-luminosity galaxy family, but instead of having low surface 
brightness like the most common early-type dwarfs, they have notorious high 
surface brightness \citep[e.g.][]{N87}.
They constitute a very rare group, as there are many galaxies classified as
compact ellipticals  \citep*[see, for instance,][]{B85}, but only 
five objects have been confirmed as such, in addition to the prototype M32. 

The known examples are all companions of larger galaxies. They are M32 
itself, a satellite of the M31 (Andromeda) spiral 
galaxy, NGC\,4486B close to M87 in the Virgo cluster 
\citep{SB84,D91}, NGC\,5846A close to the giant elliptical NGC\,5846 
(\citealp{D91}; \citealp*{Ma05}), A496cE close to the central 
cluster dominant (cD) galaxy of the cluster Abell 496 \citep{Ch07}, and two 
objects in the Abell 1689 cluster 
\citep[CG$_{A1689,1}$ and CG$_{A1689,2}$,][]{M05}. It should be noted that
object CG$_{A1689,1}$ from \citep{M05}, has a quite deviating radial velocity 
with respect to its closest projected giant elliptical.

In particular, M32 presents the following characteristics:    
(a) it is a satellite of a larger galaxy; (b) the brightness profile  
cannot be accurately fitted with a single S\'ersic law 
\citep*{G02,choi02};
(c) high surface brightness in comparison with ellipticals of the same 
luminosity \citep{N87} and correspondingly a small size \citep{G02,choi02}; 
(d) a radial change in both age and metallicity, leading however to a 
quite flat colour profile, its stellar population
being younger and more metal-rich at the centre \citep{R05}. The other 
confirmed cE galaxies show similar properties with respect to surface 
brightness, compactness and projected location close to brighter galaxies. 
However, there might be some differences in their colour gradients and 
brightness profiles (see, for example, \citealp{L96} and \citealp{F06}, 
regarding NGC\,4486B) as well as in metal content and age 
\citep[e.g.][]{SB06,Ch07}.

Some effort has been made to find more examples of these 
very rare objects in nearby clusters or groups, like Fornax \citep{D01} and 
Leo \citep{ZB98}. However, all the examined candidates have been rejected as 
cE galaxies. As the questions related to their origin and their role in the 
framework of galaxy evolution remain still open, it would be of great interest 
to find more objects and to study them in relation to their environment.

The Antlia cluster of galaxies is the third nearest well populated galaxy 
cluster after Virgo and Fornax. 
It exhibits a complex structure consisting of several subgroups, the
most conspicuous ones being dominated by the giant elliptical galaxies 
NGC\,3258 and NGC\,3268. X-ray observations showed extended emissions around 
both subgroups \citep*{ped97,nak00}. These emissions are concentrated towards 
the dominant galaxies, but extensions elongated in the direction to the other 
subgroup are also present, suggesting an ongoing merger. 
\citet*{dir03} and \citet*{B08} have shown that the globular cluster systems 
around NGC\,3258 and NGC\,3268 are elongated in the same direction as a 
connecting line between the two galaxies, resembling the X-ray results.

The photographic survey of \cite{FS90} was the first and last major effort
devoted to study the galaxy population of the Antlia cluster. They
identified, by visual inspection, 375 galaxies that
are listed in their Antlia Group Catalogue (hereafter labelled by the acronym
FS90 plus the catalogue number). It gives, among other data, a membership
status (1. definite member, 2. likely member, 3. possible member) and a
morphological type for each galaxy. FS90 classified 10 objects as E(M32?) 
or S0(M32?), and 1 object as d:E(M32?),N. 

Given the lack of an extensive analysis of Antlia's galaxy population, we  
initiated the Antlia Cluster Project with the aim at performing the first 
CCD-photometric and spectroscopic study of this cluster. In the first paper of 
this long-term project \citep[][hereafter Paper I]{SC08}, we presented  
photometric properties of the early-type galaxy population. Among 
our results it was found that early-type members define a tight 
colour-magnitude relation, spanning 9 mag from giant ellipticals to dwarf 
galaxies, without 
a perceptible change of slope. This slope is similar to those found in other 
clusters like Virgo, Fornax, Perseus and Coma, which are dynamically 
different to Antlia, and it is also consistent with that displayed by 
the so called `blue tilt' of metal-poor globular clusters in NGC\,4486. 

Among the Paper I sample, there were four galaxies 
classified as cE candidates by \citet{FS90}. Two of these
objects are spectroscopically confirmed Antlia members that seem to be normal 
low-luminosity early-type galaxies. The other two are each one close 
in projection to either one of the dominant galaxies, and are separated from 
the locus of the early-type members in the luminosity-mean effective surface 
brightness diagram. They display some characteristics of cE galaxies, 
but radial velocities are not available.

In this paper, the second of the Antlia Cluster Project, we present a 
photometric analysis of four FS90 cE candidates (namely FS90\,110, 
FS90\,165, FS90\,192, and FS90\,208) located in the central region of the
Antlia cluster. We aim at obtaining photometric evidence in
favour of, or against to, these objects being genuine cE galaxies. 
Given the small number of cE galaxies known to date, any additional 
members of this class could give important clues on their evolutionary path.
The central question is probably whether they have been dynamically 
transformed by the interaction with a massive galaxy.

Throughout this paper we will adopt $(m-M)=32.73$ as the Antlia distance 
modulus \citep{dir03}, which corresponds to an Antlia distance of 35.2 Mpc. At 
this distance 1 arcsec subtends 170 pc. 
The paper is organized as follows. In Section\,\ref{data} we give information 
about our photometric data. In Section\,\ref{analysis} and \ref{HST} we 
present the analysis of the data, and in Section\,\ref{conclusions}, a 
discussion and our conclusions.

\section{Data}
\label{data}
\subsection{Observations}
\label{observations}

\begin{figure}
\center
\includegraphics[width=85mm]{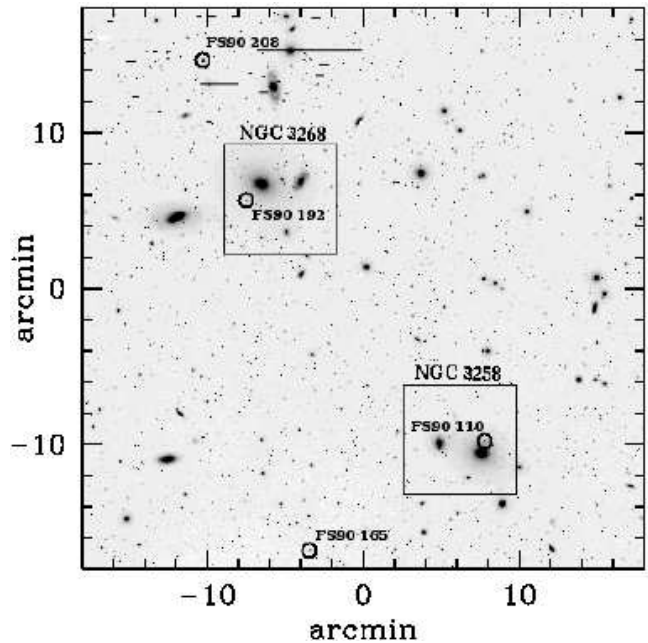}\\
\caption{$R$ image of the MOSAIC field. We show the positions of the two 
FORS1 fields centred on NGC\,3258 and NGC\,3268, respectively. The open 
circles indicate the locations of the FS90 cE candidates in the 
central region of Antlia. At the adopted Antlia distance, 
1 arcmin$=$10.2 kpc. North is up and east to the left.}
\label{mosaic}
\end{figure}

The data set comprises Kron-Cousins $R$ and Washington $C$ \citep{C76} 
images obtained with the MOSAIC camera (8 CCDs mosaic 
imager) mounted at the prime focus of the 4-m Blanco telescope at the Cerro 
Tololo Inter--American Observatory (CTIO), during 2002 April 4--5. These 
images cover the central part of the Antlia cluster (Fig.\,\ref{mosaic}) and 
were used for Paper I. One pixel of the MOSAIC 
subtends 0.27 arcsec on the sky, which results in a field of view of 
$36 \times 36$ arcmin$^2$ (i.e. about $370 \times 370$ kpc$^2$ at the Antlia 
distance). The seeing on the $R$ image is 1 arcsec and on the $C$ image 1.1 
arcsec. We refer to \citet{dir03} for more details.  

We selected the Kron-Cousins $R$ filter instead of the original Washington 
$T_1$ due to its better transmission at all wavelengths (\citet{gei96}). 
$R$ and $T_1$ magnitudes are very similar, with just a very small colour term
and zero-point difference ($R - T_1 \approx -0.02$).
We have  transformed $R$ magnitudes into $T_1$ magnitudes by the calibration 
given in \citet{dir03}. 

As a supplement to the MOSAIC observations, we use Bessel $V$ and $I$ images 
for two fields centred on NGC\,3258 and NGC\,3268, respectively, 
that have been obtained during 2003 March 27-28 with FORS1 at the 8-m VLT UT1 
(Antu) telescope (Cerro Paranal, Chile). These images cover two FS90 cE 
candidates (FS90\,110 and FS90\,192), are deeper and have 
higher resolution than those from CTIO. One pixel of this camera subtends 0.2 
arcsec on the sky, giving a field of view of 6.8 $\times$ 6.8 arcmin$^2$, 
i.e. 60 $\times$ 60 kpc$^{2}$ at the Antlia distance. The seeing on the 
$V$ image is 0.53 arcsec for the NGC\,3258 field, and 0.54 arcsec for 
the NGC\,3268 one. We refer to \citet{B08} for more details.

In addition, we have combined four F814W band images of 570 sec each, and four
F435W band frames of 1340 sec each, obtained from the ACS-HST archive and 
centred on NGC\,3258, as well as four images in the same bands and with the 
same exposure times, centred on NGC\,3268 (Proposal ID: 9427, PI: W. 
Harris). The FS90 cE candidates covered by 
these frames are, once more, FS90\,110 and FS90\,192. 
In order to make a comparison between these two candidates with a confirmed cE 
galaxy, we have also combined two images in the F850LP band of 560 sec each of 
NGC\,4486B, obtained from the same archive (Proposal ID: 9401, PI: P. 
C\^ot\'e). 

All ACS images were processed with the standard 
calibration pipeline (CALACS+multidrizzle), including bias, 
dark and flat-fielding corrections. The HST images have a FWHM of 0.1-0.12 
arcsec, one pixel of the ACS camera subtends 0.05 arcsec on the sky, and its 
field of view is 202 $\times$ 202 arcsec$^2$, i.e. 34.3 $\times$ 34.3 kpc$^2$ 
at the Antlia distance. 
Following \citet{S05}, the F850LP was calibrated to the $z$ band. Additionally,
and adopting the transformation for $(B-I)>1$, the F814W filter was calibrated 
in Cousins $I$ band magnitudes, and the F435W filter in Johnson $B$ band 
magnitudes.

\subsection{Photometry}
\label{photometry}
 We calculated total $T_1$ magnitudes and $(C-T_1)$ colours 
 by numerically 
integrating the observed brightness profiles of our FS90 cE candidates.  
To obtain these profiles we applied the task ELLIPSE within IRAF 
\citep{J87}. For each galaxy, we used the $R$ MOSAIC image  
to obtain the elliptical apertures to be used also for the $C$ image. 
These apertures were determined allowing their centres, ellipticities and 
position angles to vary freely, until their fits became unstable as a 
consequence of the low surface brightness. At this point, the three parameters 
were fixed in order to extend the isophotal fit as much as possible towards
larger radii. The sky level was determined following the procedure 
described in Paper I.

The central regions of FS90\,110, FS90\,165 and FS90\,208 were overexposed
on the long-exposure images. To include them,  
we worked with both long- and short-exposure frames following the procedure
described in Paper I for bright galaxies with overexposed centres (see also
Sect.\,\ref{profiles}). 

FS90\,110 and FS90\,192 are projected close to NGC\,3258 and NGC\,3268, 
respectively. As a consequence, they are embedded within the light of their 
bright companions. To obtain their brightness profiles, we constructed 
2-dimensional models  of both dominant galaxies, and subtracted 
them from the original images. To build these models, we used the task BMODEL
within IRAF, which creates a 2-dimensional image file containing a 
noiseless photometric model of a source image. The models are based on 
the isophotal analyses previously performed with ELLIPSE.  
 We chose a spline function for the interpolation 
which is supposed to give the smallest residuals. 

In order to ensure the best quality of our models we performed two 
sequences of modelling and subtraction. The first models of NGC\,3258/3268, 
built after masking out the neighbouring bright galaxies, served to enable the 
modelling of the companion galaxies. These models are then subtracted from 
the original images, after which we obtain the final models of the dominant 
galaxies. The subtraction of these last models for NGC\,3258 and NGC\,3268 
from the original images provides the frames in which the profiles of 
FS90\,110 and FS90\,192 are measured, following the fitting procedure 
mentioned above.

The FORS $V$ images were 
used to determine the isophotal apertures to be measured also on the FORS $I$ 
frames, and single sequences of modelling and subtraction of the dominant 
galaxies were applied. The fitting procedure was similar to that 
performed on the MOSAIC images. This is also valid for the ACS images.

Table\,\ref{tabla} lists relevant information about our cE candidates.
In the first five columns  we show the FS90 number, 
positions, FS90 morphology, and $E(B-V)$ values.  
The next six columns give the photometric results obtained in the 
Washington photometric system. These are the observed total magnitudes and 
colours (not corrected by absorption or reddening) with uncertainties in 
parentheses (obtained as described in Paper I), the surface brightness of the 
limiting isophote within which the total magnitude 
has been calculated, the equivalent radius of the limiting isophote,  
the mean surface brightness within the 
effective radius, and the effective radius (i.e. the radius containing half 
of the light). The last column gives the heliocentric radial velocities, 
when available. 

\begin{table*}
\begin{minipage}{185mm}
\begin{center}
\caption{FS90 cE candidates located in our MOSAIC field of the Antlia 
cluster.} 
\label{tabla}
\begin{tabular}{@{}ccccccccr@{}c@{}lcr@{}c@{}lr@{}c@{}l@{}}
\hline
\multicolumn{1}{c}{FS90}   & \multicolumn{1}{c} {R.A.} & \multicolumn{1}{c} {Decl.}  & \multicolumn{1}{c}{FS90} &  \multicolumn{1}{c} {$E(B-V)$} &  \multicolumn{1}{c} {$T_1$} &  \multicolumn{1}{c}{$(C-T_1)$}  &  \multicolumn{1}{c} {$\mu_{_{T1}}$}&  \multicolumn{3}{c} {$r_{_{T1}}$}  &  \multicolumn{1}{c} {$\langle \mu_{\rm eff}\rangle_{_{T1}}$} &  \multicolumn{3}{c} {$r_{\rm eff_{_{T1}}}$} &\multicolumn{3}{c}{v$_{\rm r}$} \\
\multicolumn{1}{c}{ID}  & \multicolumn{1}{c} {(J2000)} & \multicolumn{1}{c} {(J2000)} & \multicolumn{1}{c}{morph.} & \multicolumn{1}{c}{\scriptsize mag} & \multicolumn{1}{c}{\scriptsize mag}& \multicolumn{1}{c}{\scriptsize mag} & \multicolumn{1}{c}{\scriptsize mag arcsec$^{-2}$} & \multicolumn{3}{c}{\scriptsize arcsec} & \multicolumn{1}{c}{\scriptsize mag arcsec$^{-2}$} & \multicolumn{3}{c}{\scriptsize arcsec}  & \multicolumn{3}{c}{\scriptsize km s$^{-1}$}\\
\hline
110  & 10:28:53.0 & -35:35:24 & E(M32?)   & 0.085 & 15.49 (0.01) & 2.06 (0.02) & 27.5 & 14&.&0 & 18.4   &  1&.&5 &&  -  & \\
165  & 10:29:46.0 & -35:42:25 & S0(M32?)  & 0.086 & 15.50 (0.01) & 2.01 (0.02) & 27.3 & 20&.&9 & 20.3   &  3&.&6  & 2605&$\pm$&80 \\
192  & 10:30:04.5 & -35:20:31 & E(M32?)   & 0.104 & 16.66 (0.01) & 2.11 (0.02) & 27.6 & 11&.&4 & 19.8   &  1&.&7 &&  -  & \\
208  & 10:30:18.7 & -35:11:49 & S0(M32?)  & 0.103 & 14.76 (0.01) & 1.94 (0.03) & 27.5 & 30&.&1 & 19.8   &  4&.&1 & 1774&$\pm$&100    \\
\hline
\end{tabular}
\end{center}
\medskip

{\it Notes.-} Coordinates have been obtained through CDS, which are calculated 
from FS90. Extinction values are from \citet{S98}. $\mu_{_{T_1}}$ corresponds 
to the surface brightness of the outermost isophote within which integrated 
magnitudes and colours were measured. $r_{_{T_1}}$ is the equivalent radius 
($r=\sqrt{a\cdot b}=a\cdot \sqrt{1-\epsilon}$) of that isophote. 
$\langle \mu_{\rm eff}\rangle$ is obtained within $r_{\rm eff}$, the radius 
that contains half of the light. All these values were obtained from ELLIPSE. 
The radial velocities are from 6dF.
\end{minipage}
\end{table*}

\section{Analysis of MOSAIC data}
\label{analysis}

\subsection{Morphology and spatial location}
\label{morphology}

\begin{figure*}
\includegraphics[width=51mm]{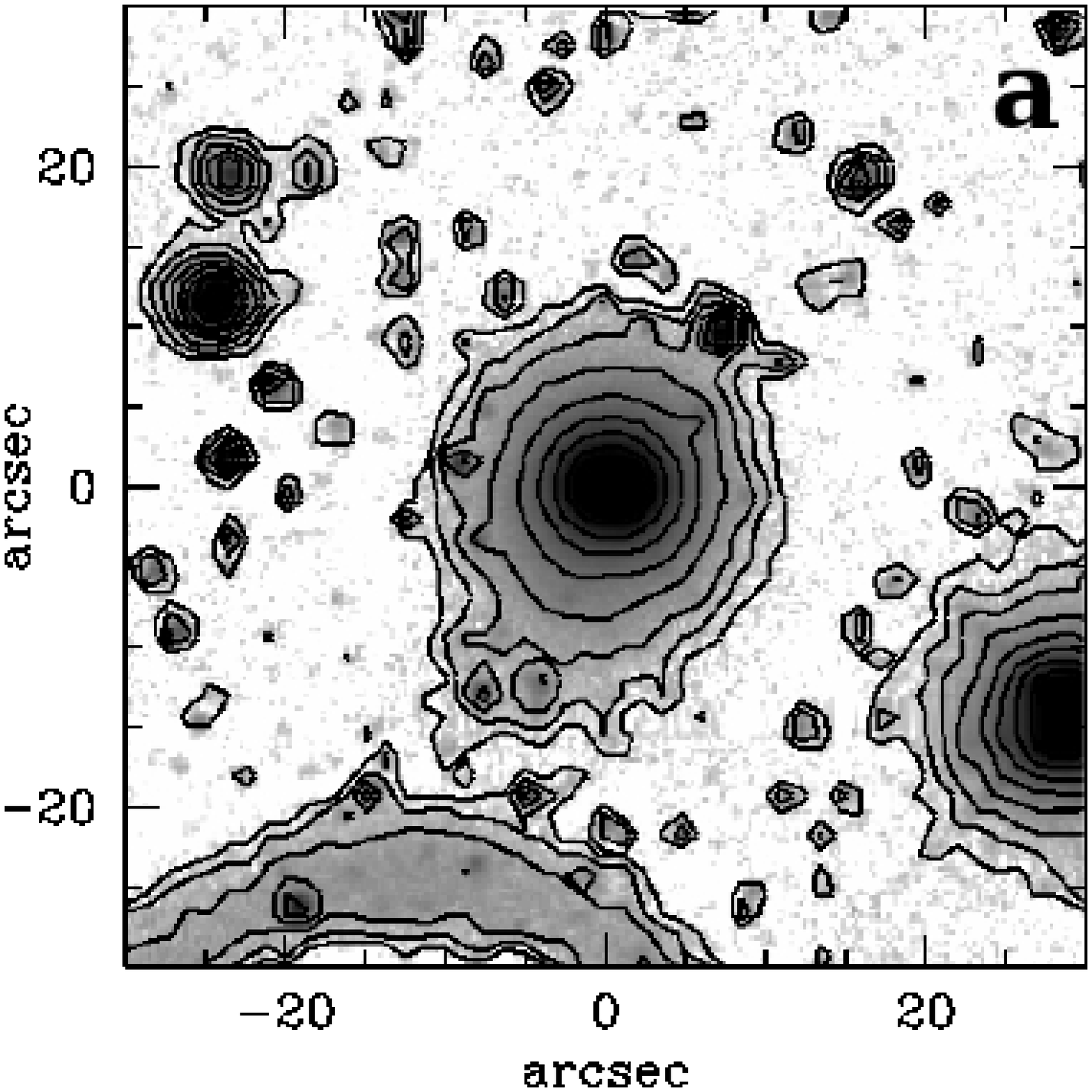}
\includegraphics[scale=0.256]{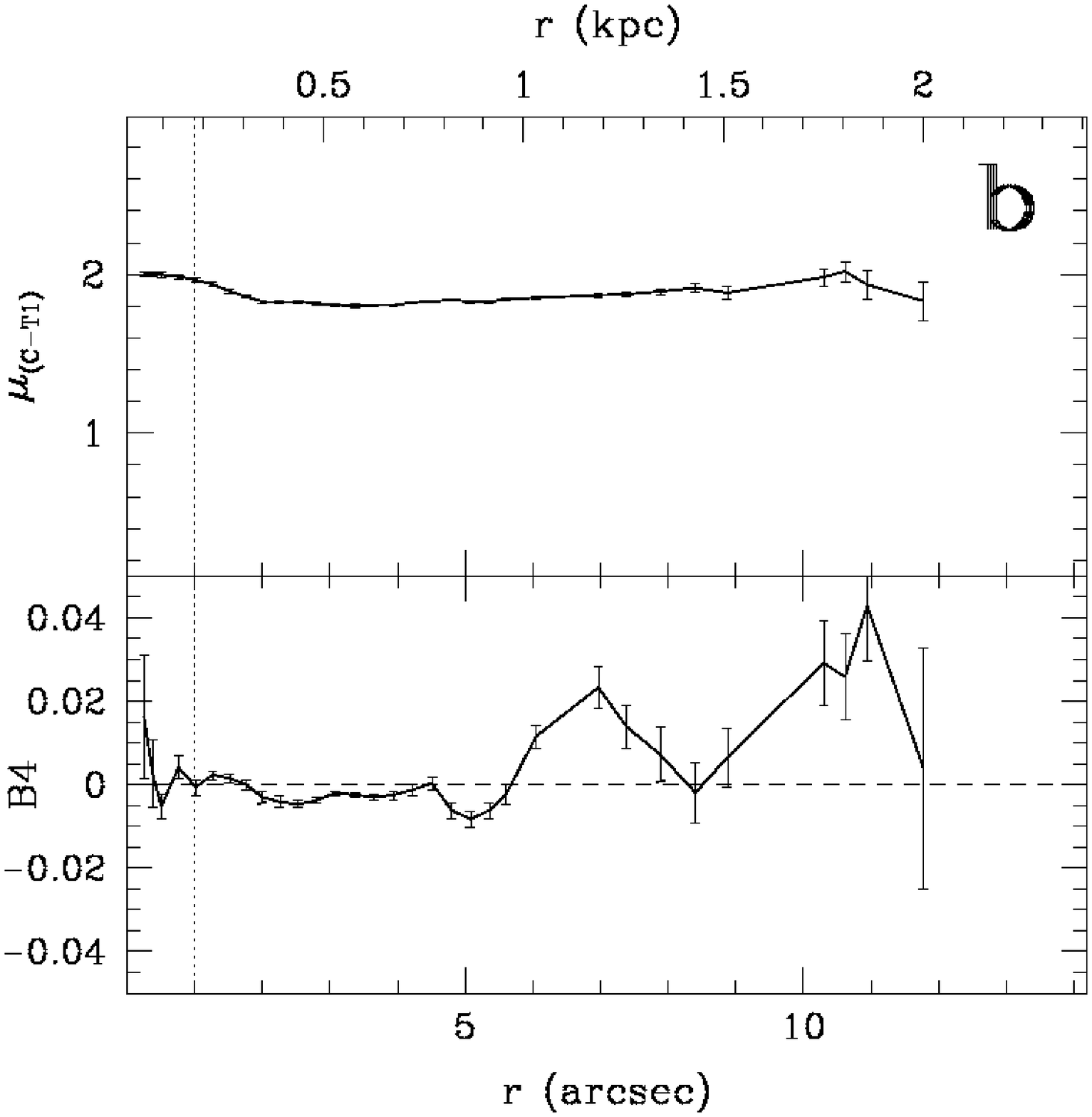}
\includegraphics[scale=0.256]{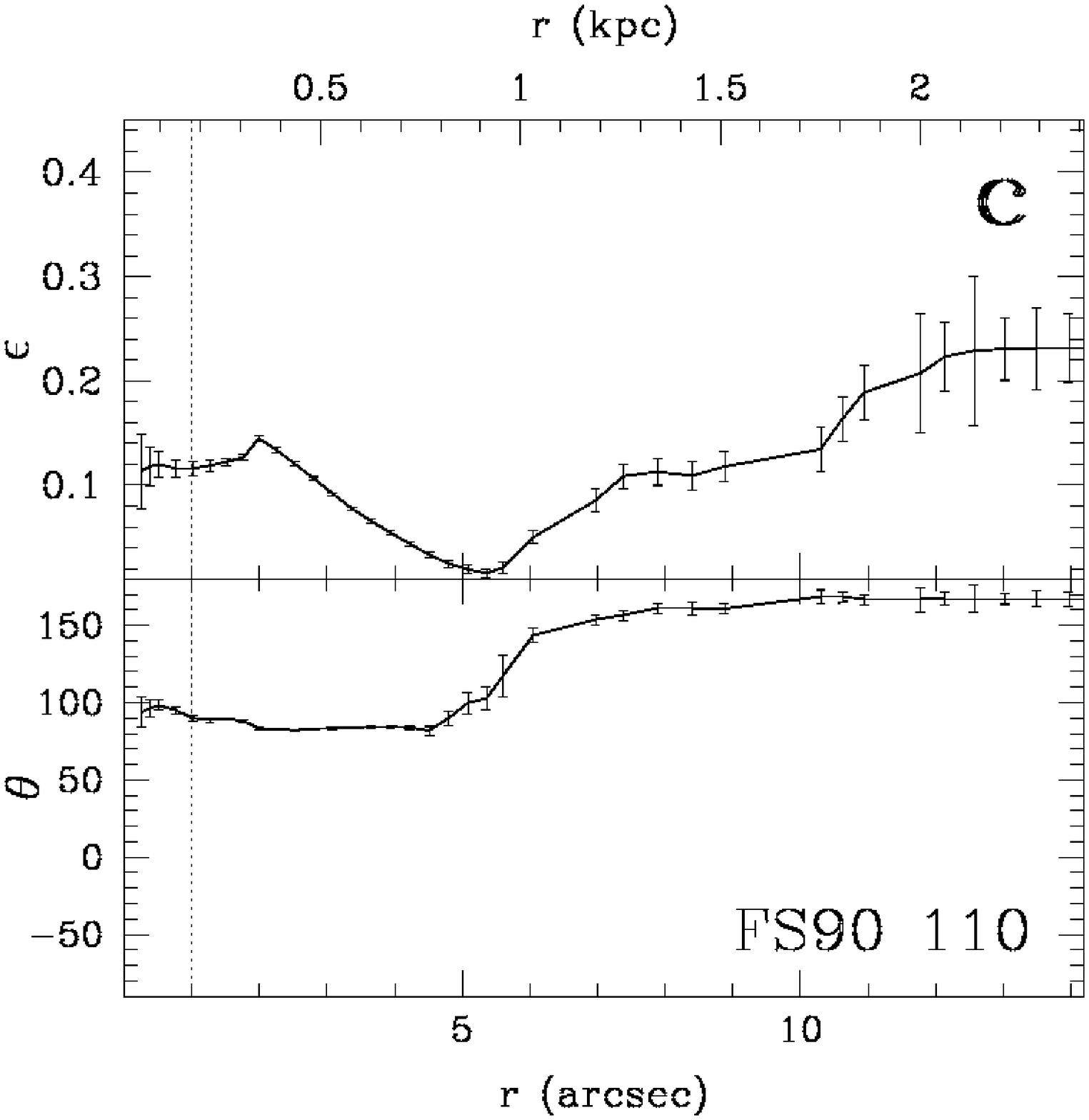}\\
\includegraphics[width=51mm]{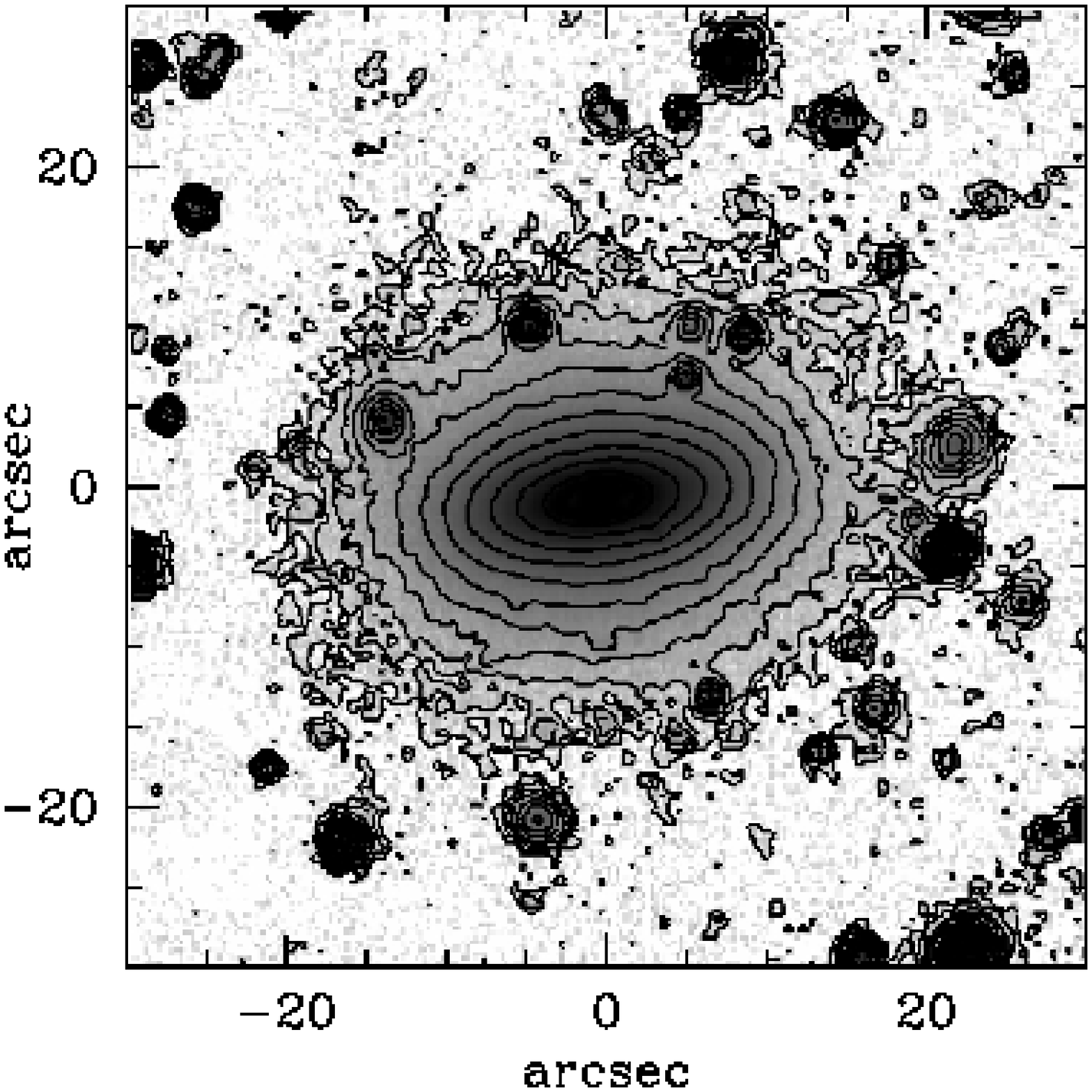}
\includegraphics[scale=0.256]{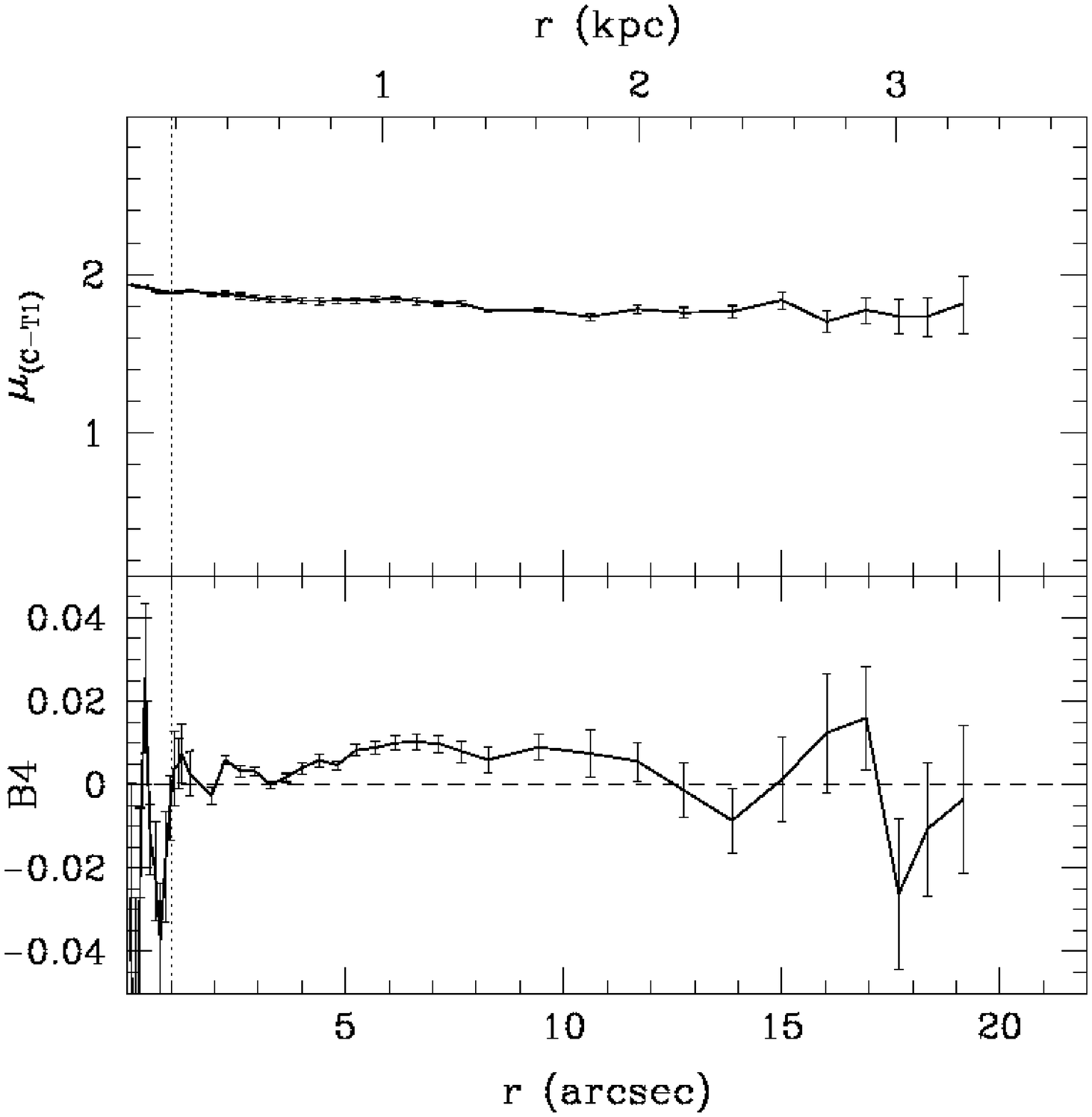}
\includegraphics[scale=0.256]{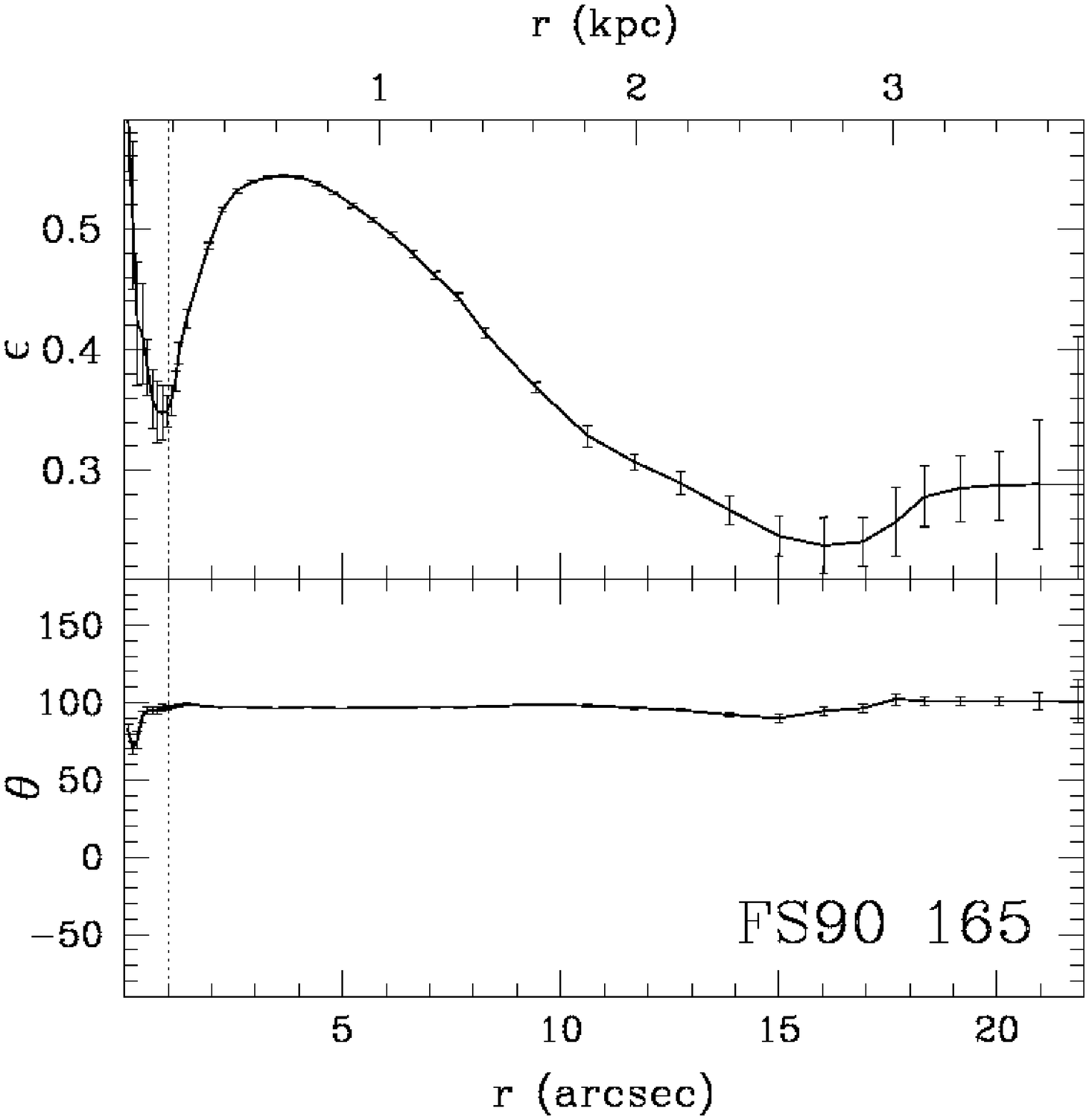}\\
\includegraphics[width=51mm]{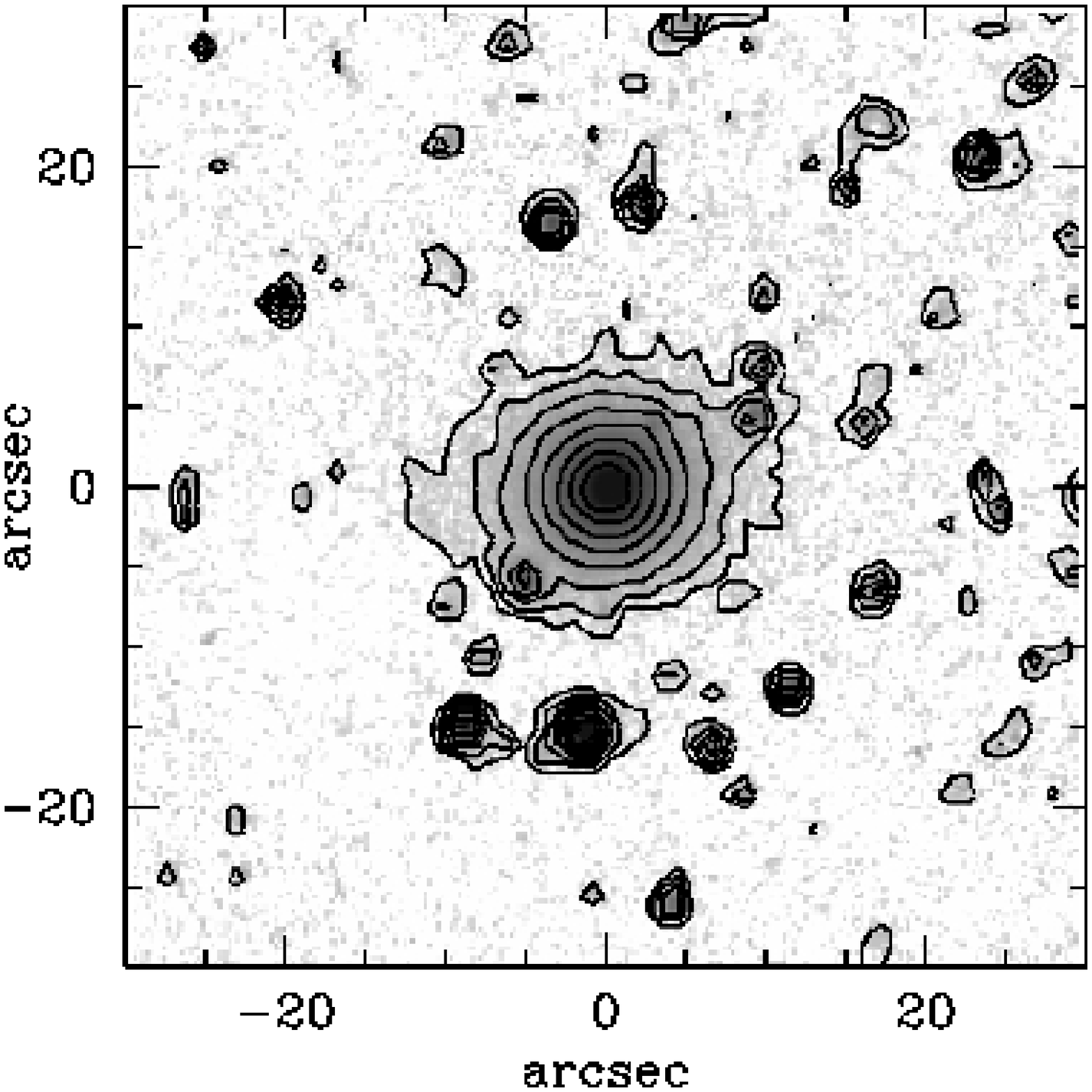}
\includegraphics[scale=0.256]{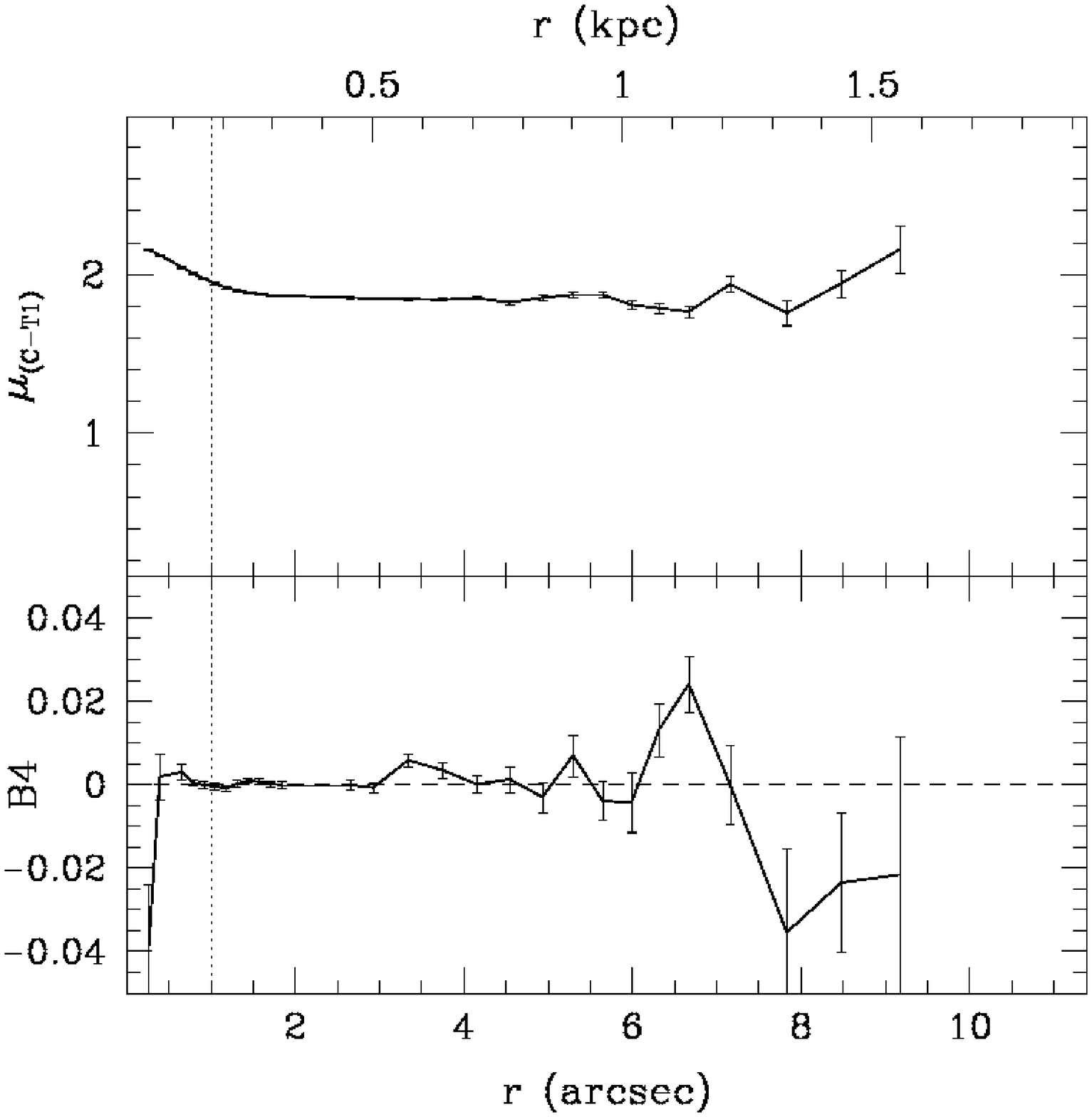}
\includegraphics[scale=0.256]{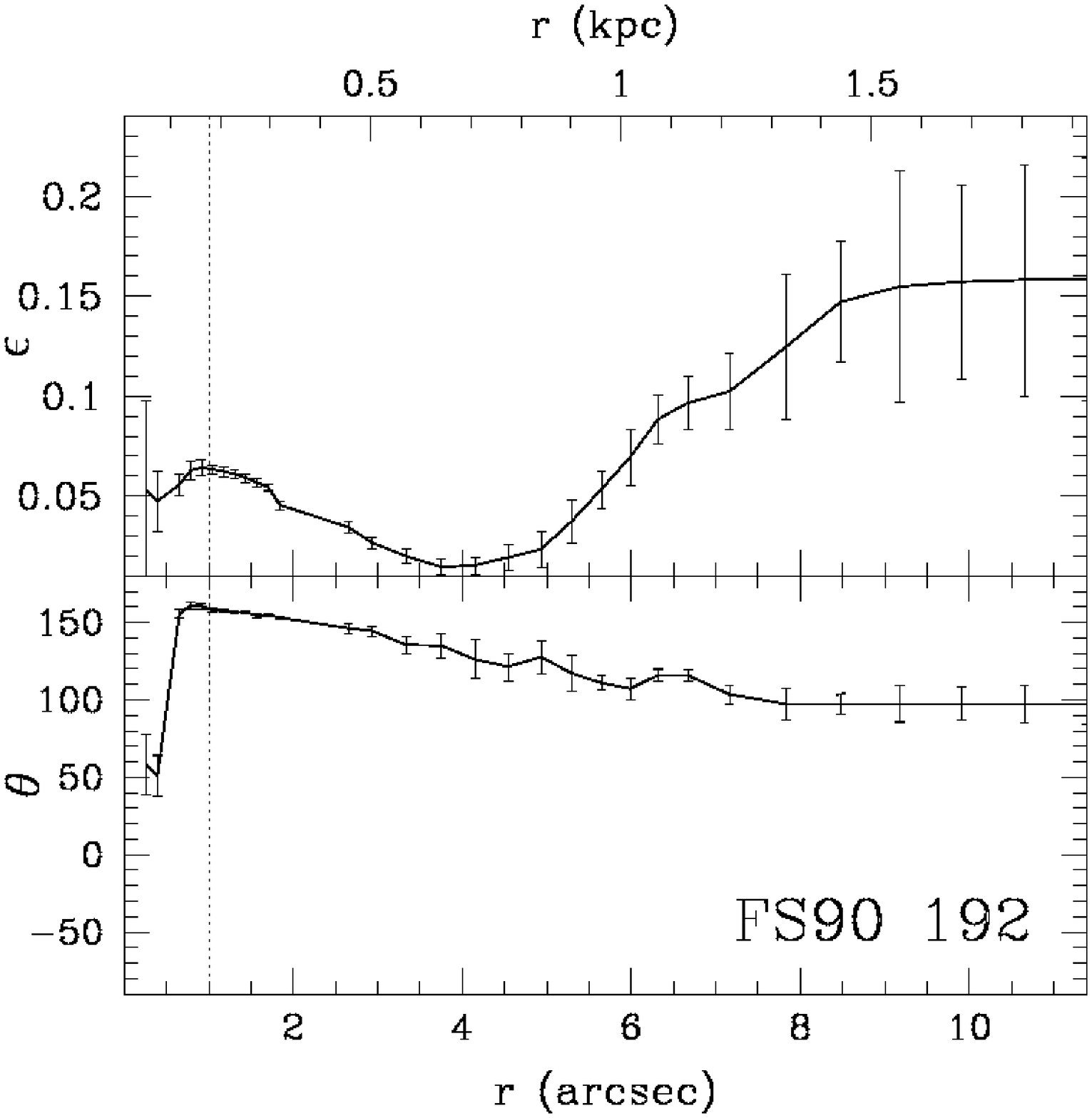}\\
\includegraphics[width=51.0mm]{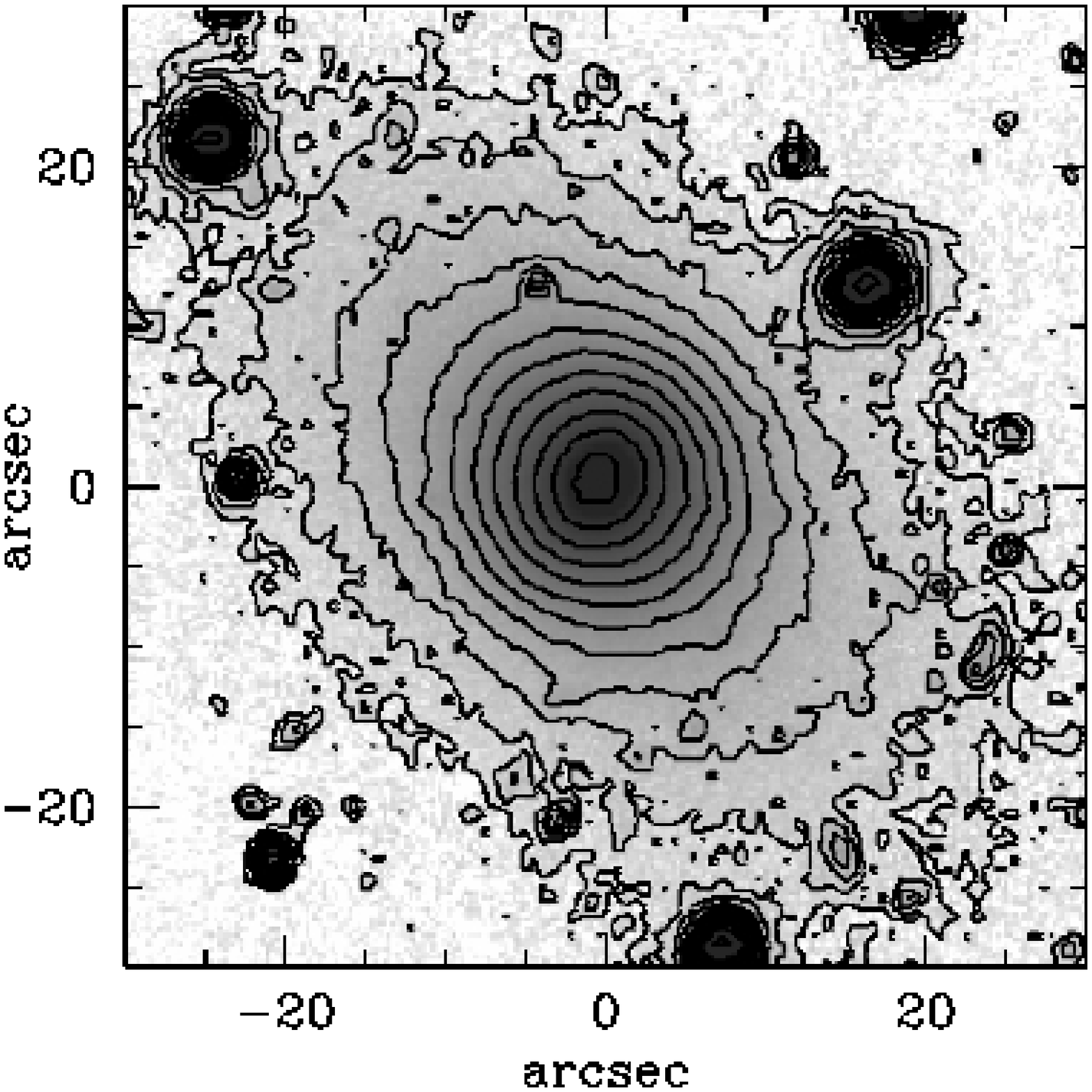}
\includegraphics[scale=0.256]{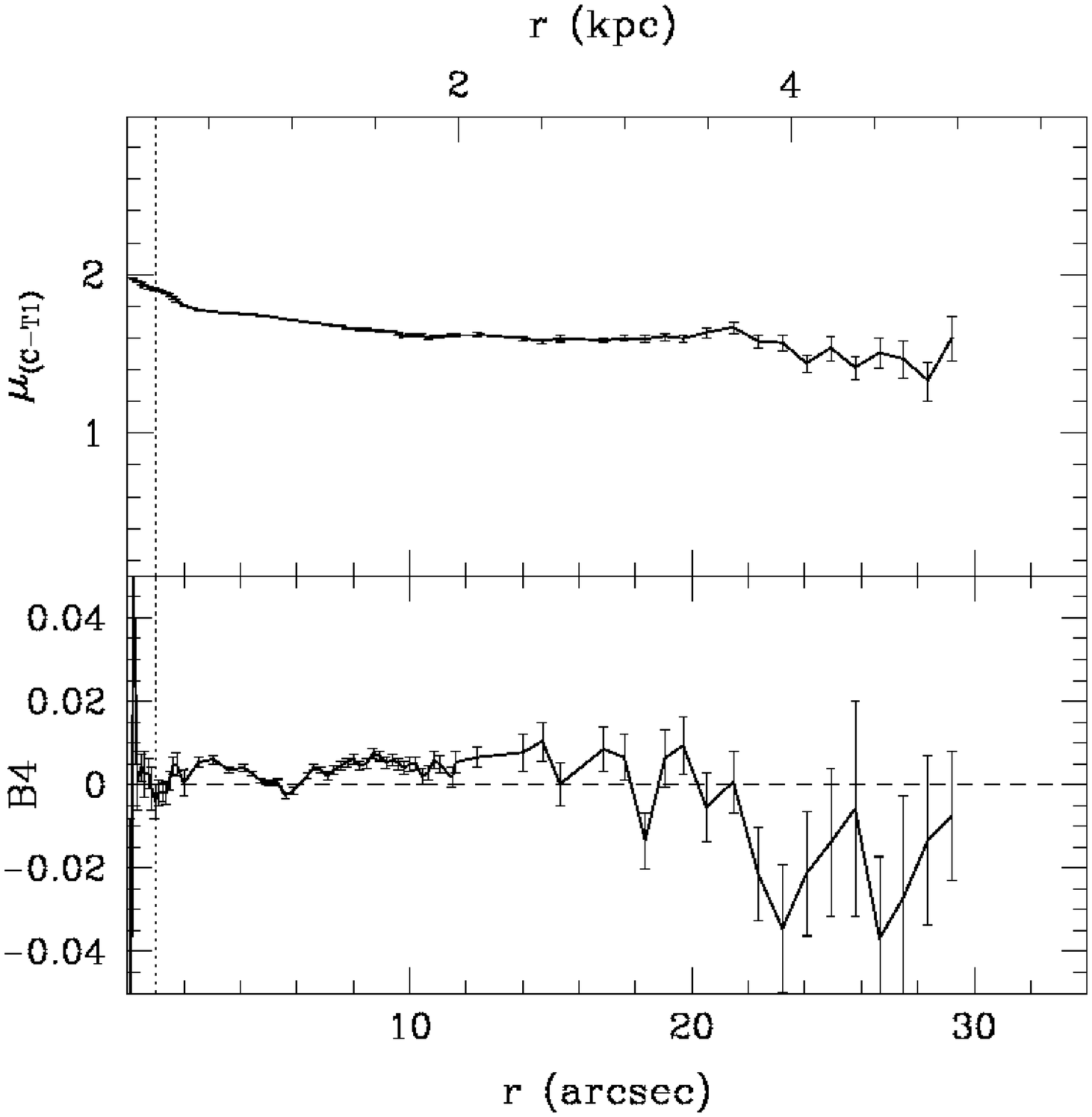}
\includegraphics[scale=0.256]{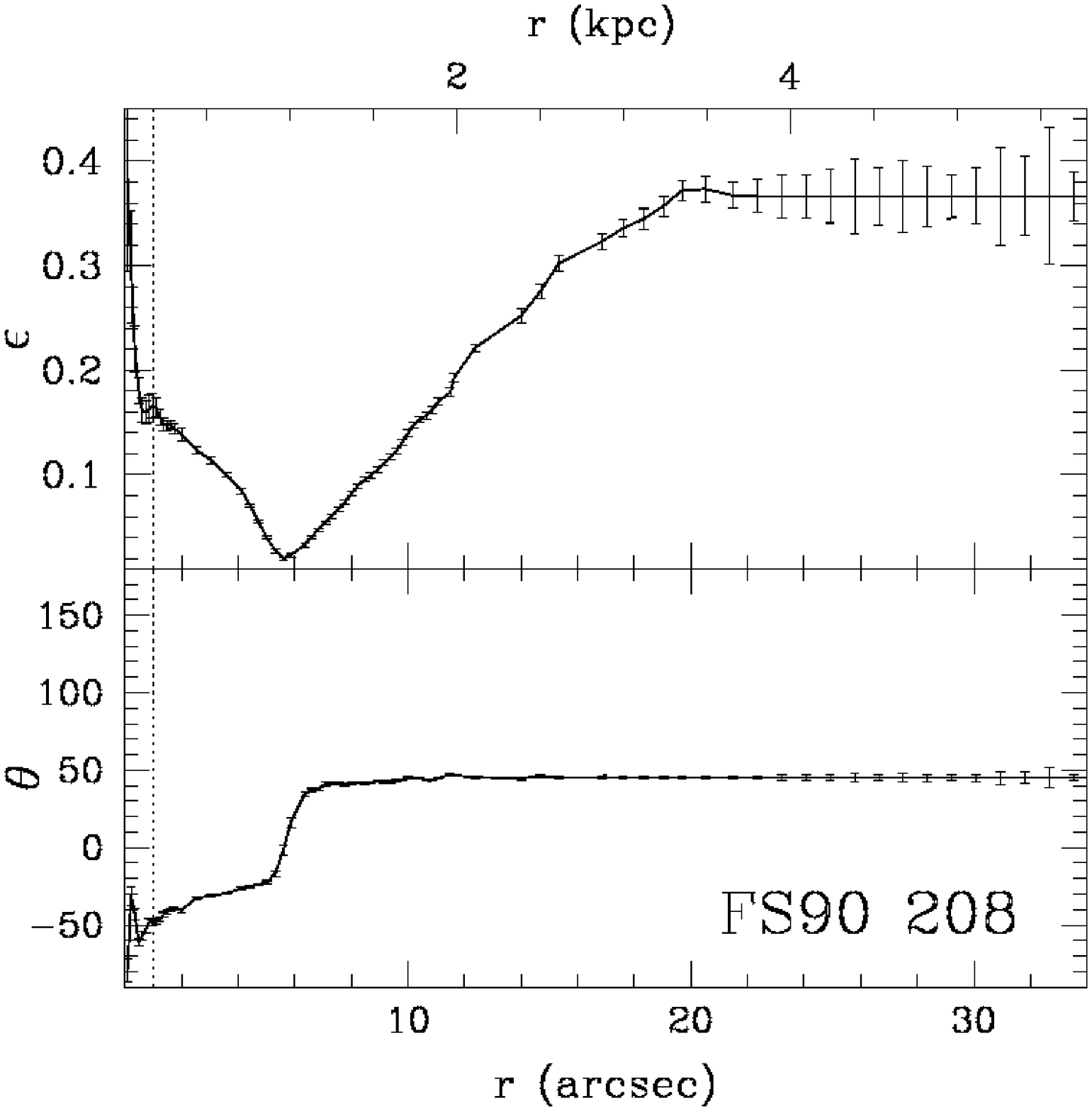}
\caption{FS90 cE candidates located in the central region of the 
Antlia cluster. {\it From top to bottom:} FS90\,110, FS90\,165, FS90\,192 and 
FS90\,208. {\it From left to right:} (a) Brightness contour levels superimposed 
on the $R$ images of the galaxies. At the adopted Antlia distance, 1 arcsec  
$\simeq$ 170 pc. North is up and east to the left. (b) $(C-T_1)$ 
colour profiles reddening corrected ({\it top}), and ELLIPSE B4 index 
({\it bottom}) against equivalent radius. (c) Ellipticity ($\epsilon$, 
{\it top}) and position angle ($\theta$, {\it bottom}) against equivalent 
radius. Positive angles are measured from north to east. The vertical dotted 
lines in (b) and (c) show the region of seeing influence. The equivalent 
radius scale displayed at the top of (b) and (c) panels, was obtained with the 
assumed Antlia distance.}
\label{Imagenes}
\end{figure*}

The {\it a}-panels of Fig.\,\ref{Imagenes} show the morphologies of 
the FS90 cE candidates in the MOSAIC R-frames. Each side is 1 arcmin 
(10.2 kpc at the adopted Antlia distance). In the case of FS90\,110, 
a model of NGC 3258 has been subtracted.

FS90\,110 is located at a radial distance of 47 arcsec ($\sim$ 8 kpc at the 
Antlia distance) from NGC\,3258 to the North. There is no radial velocity 
available for this object. FS90\,110 was catalogued by FS90 as a `possible' 
(i.e. status 3) 
Antlia member with an E(M32?) morphology. It resembles a cE galaxy 
(see Fig.\,\ref{Imagenes}), as it is round and small in projected size 
($r_{_{T_1}}\sim$ 2.4 kpc). It is the closest galaxy in projection to either 
one of the dominant Antlia galaxies and, interestingly, its $R$ image shows 
an elongation of the outer isophotes towards its bright companion (see 
Fig.\,\ref{Imagenes}). This low surface brightness feature is seen on the $C$ 
image as well.

FS90\,165 is located at 12.2 arcmin ($\sim$ 124.4 kpc at the Antlia distance)
from NGC\,3258 towards the SE, and is a spectroscopically confirmed Antlia 
member. It was catalogued by FS90 as a `likely' (i.e. status 2) member of 
Antlia with an S0(M32?) morphology. FS90\,165 is one of the smallest
S0 galaxies in the central region of Antlia. It is fainter than $T_1=14$ mag, 
the limiting magnitude that separates dwarfs from bright early-type galaxies 
in our Paper I sample. 

FS90\,192 is located at 1.4 arcmin ($\sim$ 14.2 kpc at the Antlia distance) 
from NGC\,3268 to the SE. No radial velocity is available. It is 
catalogued as a `possible' (i.e. status 3) Antlia member displaying an E(M32?) 
morphology. It looks round and compact like FS90\,110, albeit smaller 
($r_{_{T_1}}\sim$ 1.9 kpc at the Antlia distance, see Table\,\ref{tabla}). It 
is more distant in projection from NGC\,3268 than FS90\,110 is from NGC\,3258, 
and there is no visible distortion in its outer isophotes.

FS90\,208 is a confirmed Antlia member located at 8.5 arcmin  
($\sim$ 86.7 kpc at the adopted Antlia distance) from NGC\,3268 to the NE. It 
is considered to be  a 
S0(M32?) `definite' (i.e. status 1) member of Antlia by FS90. It displays a 
round morphology in its central region and gets elongated outwards. It is 
the largest FS90 cE candidate in the central region of Antlia 
(see Table\,\ref{tabla}).

\subsection{Brightness and colour profiles}
\label{profiles}

\begin{figure*}
\includegraphics[width=86mm]{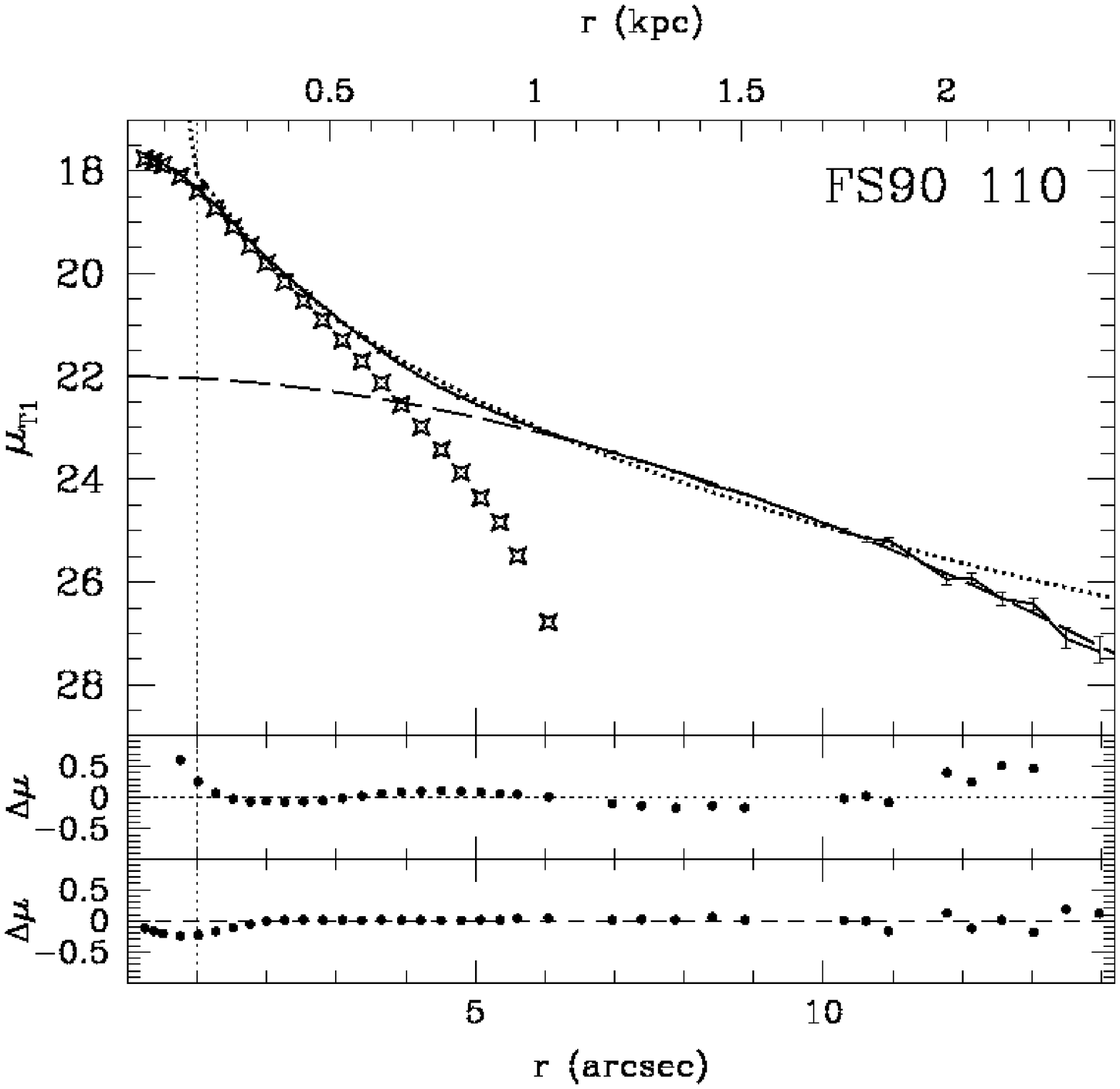}
\includegraphics[width=86mm]{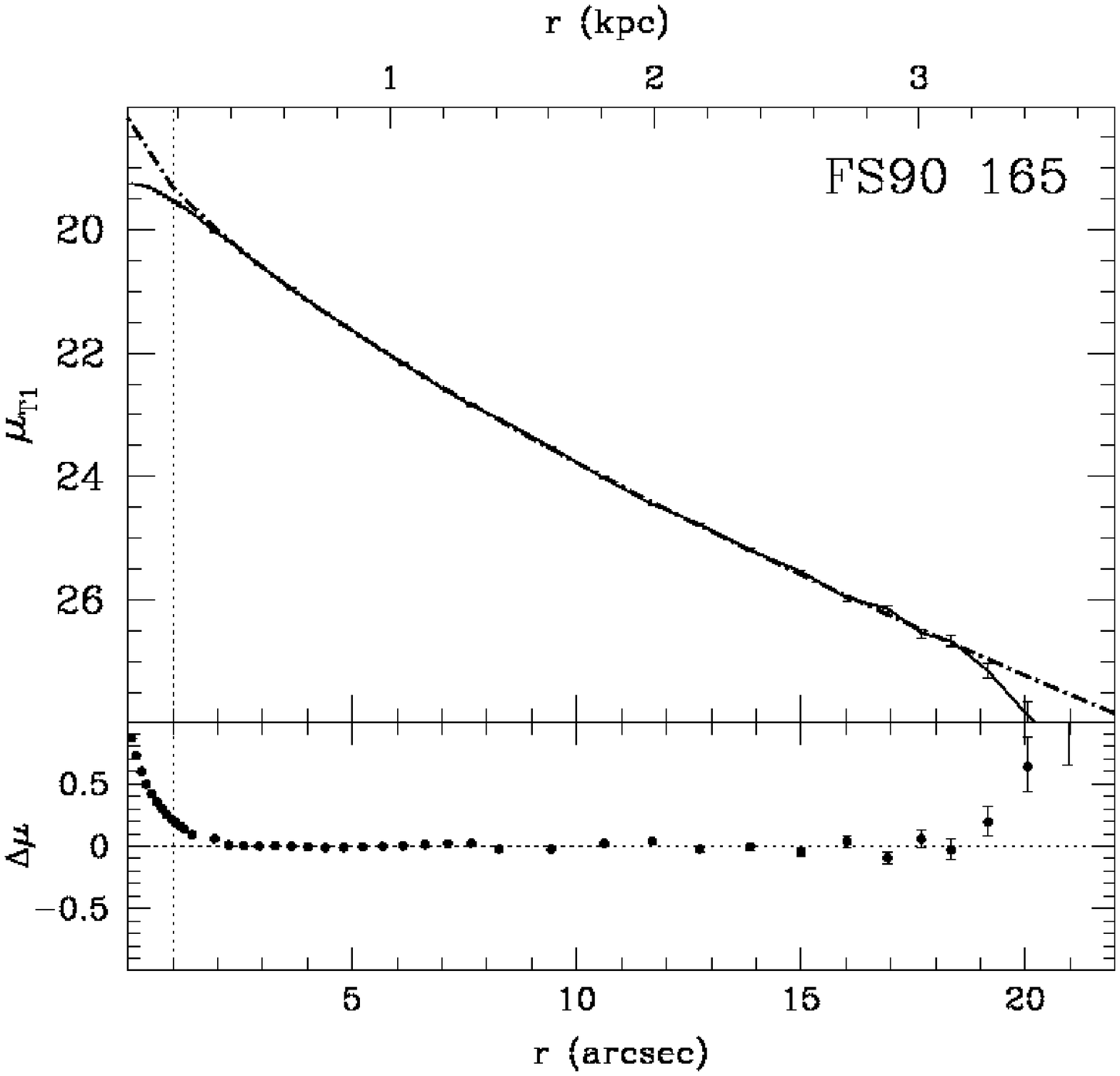}\\
\includegraphics[width=86mm]{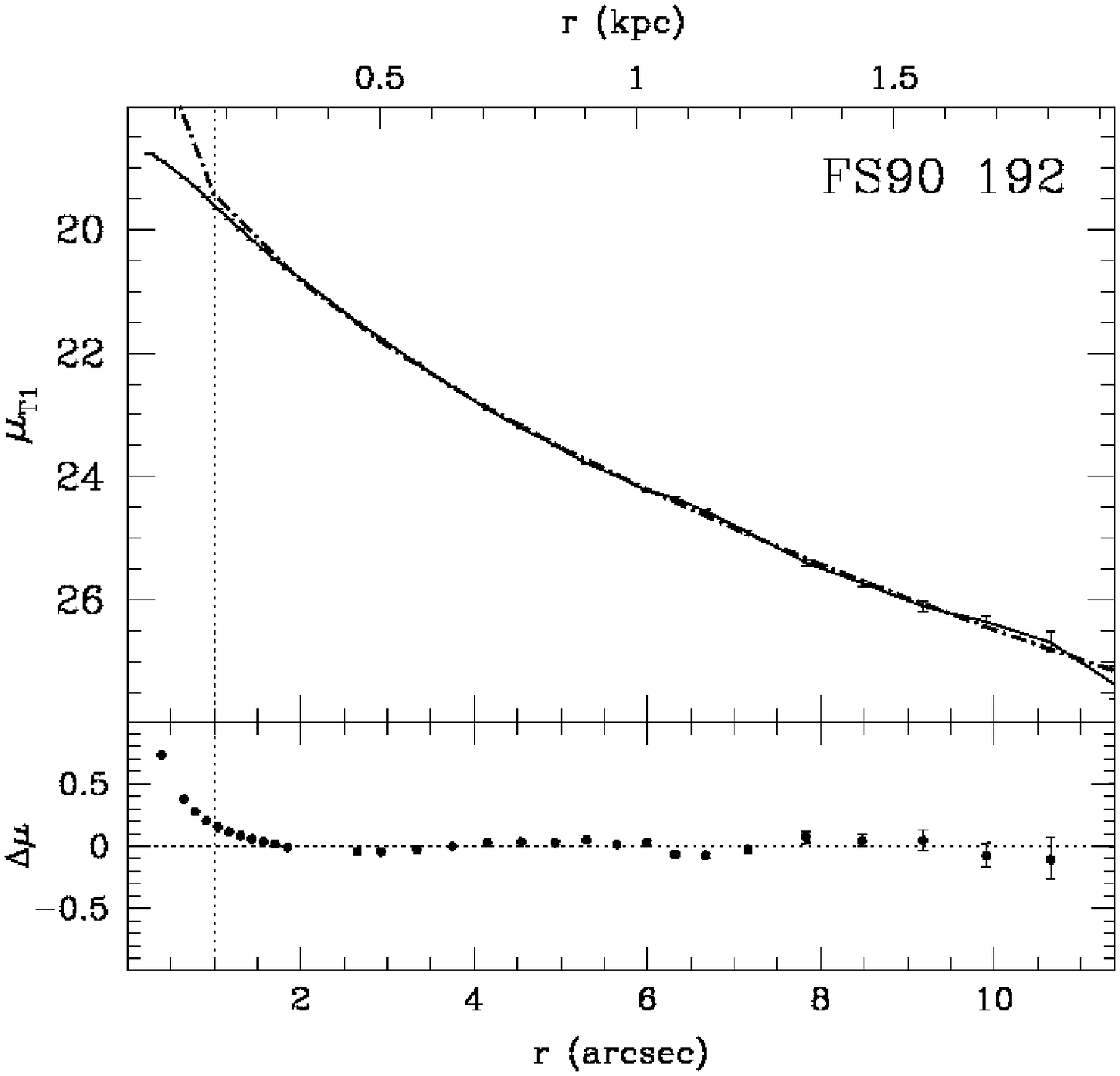}
\includegraphics[width=86mm]{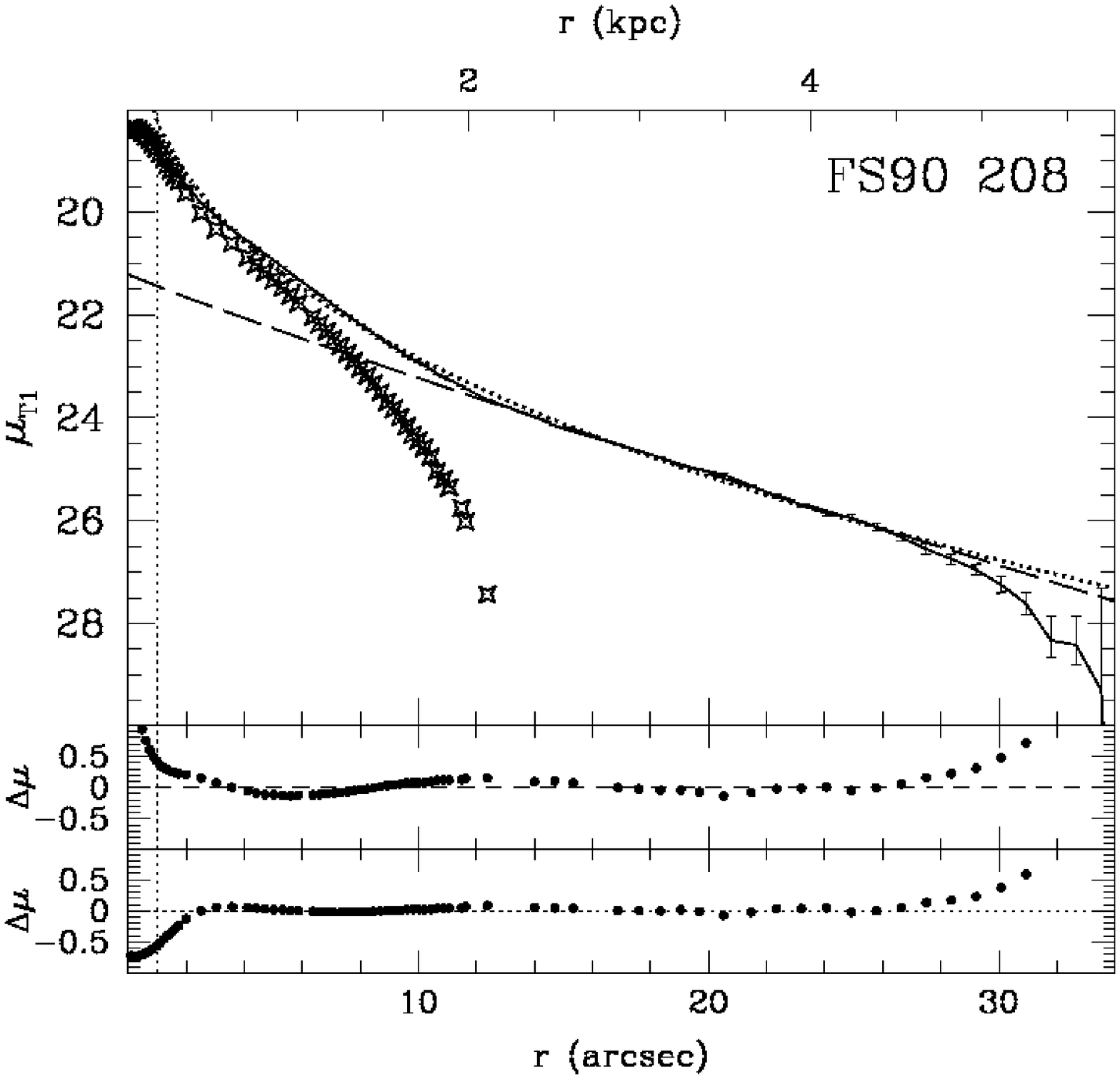}
\caption{Fits of the S\'ersic law to the absorption corrected $T_1$ 
brightness profiles of the FS90 cE candidates located in the central region of
Antlia. The residuals $\Delta \mu=\mu(obs)-\mu(fit)$ are shown in the lower 
panels. For FS90\,110 and FS90\,208, which display two
component profiles, the upper residuals correspond to the fit of a 
single S\'ersic law (dotted curve), and the lower ones to a two component  
fit. For clarity, in these cases we only show the S\'ersic fit for the 
outer component (dashed line), which is subtracted to show the inner one 
(open symbols). The equivalent radius scale at the top of the four panels 
was obtained from the adopted Antlia distance (1 arcsec $\simeq$ 170 pc).}
\label{ajustes_Sersic}
\end{figure*}

As mentioned above, panels {\it a} of 
Fig.\ref{Imagenes} show contour plots superimposed on the $R$ images of the 
FS90 cE candidates. We also present their $(C-T_1)$ colour profiles 
and the ELLIPSE B4 index vs. equivalent radius (panels {\it b}), as well 
as the ellipticity ($\epsilon$) and position angle ($\theta$) vs. equivalent 
radius (panels {\it c}). Positive angles are measured from north to east. 

In Fig.\ref{ajustes_Sersic} 
we show the brightness profiles for these four galaxies.
From these plots it can be seen that, if we exclude the region affected by 
seeing ($r<1$ arcsec), FS90\,165 and FS90\,192 show what seems
to be a one component profile with smooth variations in ellipticity and 
position angle. On the contrary, the brightness profiles of FS90\,110 and 
FS90\,208 seem to present two components in agreement with the strong 
changes both in ellipticity and orientation displayed by the fitted apertures.

To test if single component models provide good fits to the brightness 
profiles of the FS90 cE candidates, we used the S\'ersic law \citep{S68}:
\begin{equation}
\mu(r)=\mu_0+1.0857~\big(\frac{r}{r_0}\big)^N .
\label{Sersic1}
\end{equation}
where $\mu_0$ designates the central ($r=0$) surface brightness, $r_0$ is the 
scalelength of the profile, and $N$ is the S\'ersic index. 
Because of its simpler mathematical dependence on the free parameters, 
we decided to use the above formula instead of:
\begin{equation}
\mu(r)=\mu_{\rm eff}+1.0857~b_n~\big[\big(\frac{r}{r_{\rm eff}}\big)^{1/n}-1\big]
\label{Sersic2}
\end{equation}
where $b_n\approx1.9992n-0.3271$ for $0.5<n<10$ 
\citep[][and references therein]{G08}. There are simple relations 
between the quantities involved in both equations \citep*[e.g.][]{Mc03}:
\begin{equation}
n=1/N
\end{equation}
\begin{equation}
\mu_{\rm eff}=\mu_0+1.0857~b_n
\end{equation}
\begin{equation}
r_{\rm eff}=r_0~b_n^n
\end{equation}
which let us easily obtain the effective radius ($r_{\rm eff}$) and the 
surface brightness at this radius ($\mu_{\rm eff}$) from the parameters
of Eq.\,\ref{Sersic1}. Following the results of the numerical
simulations performed by \citet{G05}, it was decided not to correct the 
S\'ersic parameters for seeing effects (\citealp{T01a},b) unless a S\'ersic 
index n greater than 3 was obtained.

The structural parameters obtained from the fits are listed in 
Table\,\ref{ajustes_mosaic}. The profile fits and their residuals 
$\Delta \mu=\mu(obs)-\mu(fit)$ are plotted in Fig.\,\ref{ajustes_Sersic}.

\begin{table*}
\centering
\caption{Structural parameters of FS90 cE candidates, obtained
from different fits of the S\'ersic law to their absorption corrected 
$T_1$ brightness profiles. $r_{int}$ and $r_{ext}$ refer to the inner and 
outer equivalent radius considered for the fits, respectively. 
$T_1$ gives the integrated magnitude obtained from the fitted S\'ersic profile.}
\begin{tabular}{@{}lcccccccccccc}
\hline
{Object} & {Component} & $r_{int}$	 & $r_{ext}$  & {$\mu_0$}                        & {$r_0$}               & {N} & {$\mu_{\rm eff}$}                & {$r_{\rm eff}$}  & {n}  & {$T_1$}   \\ 
{}       & {} &  {\scriptsize(arcsec)}  &  {\scriptsize(arcsec)} & {\scriptsize(mag arcsec$^{-2}$)} & {\scriptsize(arcsec)} & {}  & {\scriptsize(mag arcsec$^{-2}$)} & {\scriptsize(arcsec)} &\\ 
\hline										
FS90 110 & inner & 2.3  &  4.8 & 17.73$\pm$0.02 & 1.17$\pm$0.01         & 1.23$\pm$0.01 & 19.13 & 1.43 & 0.81 & 15.61\\ 
FS90 110 & outer & 6.0  & 11.8 & 21.99$\pm$0.45 & 6.07$\pm$2.78         & 1.83$\pm$0.36 & 22.82 & 5.26 & 0.55 & 16.84\\ 
& & & & & & & & & & & \\
FS90 165 & single   & 2.2  & 13.8 & 18.19$\pm$0.06 & 0.94$\pm$0.05         & 0.69$\pm$0.01 & 20.98 & 3.69 & 1.45 & 15.25\\ 
& & & & & & & & & & & \\
FS90 192 & single   & 1.6  &  9.9 & 15.77$\pm$0.34 & 0.07$\pm$0.02         & 0.46$\pm$0.02 & 20.12 & 1.42 & 2.17 & 16.27\\ 
& & & & & & & & & & & \\
FS90 208 & inner & 2.0  &  9.8 & 19.14$\pm$0.08 & 2.92$\pm$0.29         & 1.29$\pm$0.04 & 20.44 & 3.37 & 0.77 & 15.04\\ 
FS90 208 & outer & 12.4 & 24.9 & 21.21$\pm$0.55 & 5.26$\pm$2.45         & 0.93$\pm$0.14 & 23.16 & 9.90 & 1.07 & 15.52\\ 
\hline
\end{tabular}
\label{ajustes_mosaic}
\end{table*}

\subsubsection{FS90\,110}

As it was mentioned in Sec.\,\ref{photometry}, the centre of FS90\,110 
is overexposed in the $R$ long-exposure image obtained with MOSAIC. Thus, 
we used the short-exposure one to obtain its inner
brightness profile up to an equivalent radius of $r=1.8$ arcsec.
For larger radii, we used the $R$ long-exposure frame to obtain its 
brightness profile out to $r\sim 14$ arcsec ($\mu_{T1}= 27.5$ mag 
arcsec$^{-2}$), starting with the parameters for $r=1.8$ arcsec.

The centre of the elliptical apertures, the ellipticity $\epsilon$ 
and the position angle $\theta$ were allowed to vary freely for 
the inner fit. For the outer fit, they were fixed once the 
convergence of the isophotal fit was prevented due to the low surface 
brightness of the outermost isophote.

The ellipticity shows a strong variation on a small radial scale. It
changes from $\epsilon=0.14$ at $r\sim 2$ arcsec, to $\epsilon=0.01$ at
$r=5.3$ arcsec. Then, it increases up to $\epsilon=0.22$ at $r=12.3$ arcsec and
after that it remains constant. This behaviour is followed by the position
angle as it varies from $\theta\sim 83\degr$ in the range 
$r=2 - 4.5$ arcsec, to $\theta\sim 160\degr$ at $r\sim 8$ arcsec. The B4
coefficient becomes positive (i.e. disky isophotes) in the outer region, 
at a similar radius as for which the ellipticity and the position angle 
change in a significant manner.

The outer isophotes of FS90\,110 display an elongation towards NGC\,3258
in the $R$ MOSAIC image, as it can be seen from Fig.\,\ref{Imagenes}. 
Therefore, we performed several isophotal fits to test if 
the strong variation in position angle and ellipticity could be an artifact 
due to the presence of this low surface brightness structure. We performed 
an outer fit keeping the position of the centre fixed during the whole 
process, and an outer fit masking half of the galaxy in the direction to 
NGC\,3258. In all these cases, $\theta$ and $\epsilon$ show, within the errors,
the same behaviour like in the free isophotal fit.

The colour profile of this galaxy does not show any perceptible gradient out
of the region of seeing influence.

We performed several fits of the S\'ersic law to the $T_1$ brightness
profile of FS90\,110. It was found that it is more convenient to fit
the profile with two components, as a single component fit gives systematic 
positive and negative residuals (see Fig.\,\ref{ajustes_Sersic}).
Furthermore, variations of the range in equivalent radius used to perform 
single component fits, make the S\'ersic index $N$ to evolve from convex 
profiles for larger ranges to concave ones for smaller intervals. 

Following \citet{CB01}, to get possible analytical profiles 
for the two components we first 
fitted a S\'ersic law to the outer region of the galaxy 
in the range $r=6.0 - 11.8$ arcsec (dashed line in FS90\,110 panel of
Fig.\,\ref{ajustes_Sersic}). Then, we subtracted the intensities of this
model from the observed ones in the whole profile range.
In this way we recover the inner component (open symbols in 
Fig.\,\ref{ajustes_Sersic}), which was then fitted with an 
independent S\'ersic law (not shown for clarity in 
Fig.\,\ref{ajustes_Sersic}). 

We are aware that this decomposition scheme might be not unique as it 
depends on the radial ranges selected to perform the fits, as well as on sky 
subtraction effects \citep[see, for instance,][for a discussion about the art 
of profile fitting]{CB01}. However, the validity of our approach is supported 
by the small and stable residuals shown in Fig.\,\ref{ajustes_Sersic}. 
Moreover, a fit of two coupled general S\'ersic laws to the profile of 
FS90\,110 in the range 2.3 -- 11.8 arcsec has given structural parameters 
that are in agreement, within the errors, with those given in 
Table\,\ref{ajustes_Sersic}. 

The integrated magnitudes obtained for both components from their individual
S\'ersic fits show that the inner component could be $\sim$ 3 times 
brighter than the outer one, making this galaxy a bulge-dominated system. 
It is worth noting that the radius at which the changes in ellipticity
and position angle arise, is similar to that at which the outer
component seems to begin to dominate  the brightness profile.

\subsubsection{FS90\,165 (confirmed Antlia member)}

The fit of elliptical apertures in the central region of FS90\,165
was performed on the short-exposure $R$ image, as its
centre is overexposed in the long-exposure one. The inner
profile was obtained allowing centre, ellipticity and position
angle to vary freely. The fit reaches an equivalent radius of $r\sim1.4$ 
arcsec, and from this radius outwards, we worked with the long-exposure $R$ 
image to obtain an outer profile reaching $r\sim19.2$ arcsec ($\mu_{T1}= 27.3$ 
mag arcsec$^{-2}$). 

Again, we allowed for free variation of all elliptical apertures parameters 
until the outermost (low surface brightness) regions of the galaxies were 
reached.

The ellipticity shows an almost constant value of about $\epsilon=0.53$ from 
$r\sim 2$ arcsec to $r\sim 4$ arcsec, and then displays a smooth decrease  
until $\epsilon=0.23$ at $r\sim 16$ arcsec. The position angle keeps a constant 
value of $\theta\sim 100\degr$ along the whole profile. This behaviour of 
the elliptical apertures can be seen in panel {\it a} of Fig.\,\ref{Imagenes}: 
the ellipses get more and more elongated towards the centre, while their 
orientation has no detectable variation. The B4 coefficient is positive over 
the whole equivalent radius range, showing that the isophotes of this
galaxy are disky.

Outside $r=2$ arcsec, the colour profile displays a slight blue 
gradient of $\sim 0.1$ mag, from $(C-T_1)_0= 1.87$ at $r=2.25$ arcsec, 
to $(C-T_1)_0=1.76$ at $r=13.8$ arcsec.

The brightness profile was fitted by a single S\'ersic law in the
range $r\sim 2- 14$ arcsec. The analytical profile is shallower than a 
de Vaucouleurs law, and the effective radius obtained from it, is 
in good agreement with that measured from the observed one (see Table\,
\ref{ajustes_mosaic}).

\subsubsection{FS90\,192}

As the centre of this galaxy is not overexposed in the
long-exposure frames, the fit of elliptical apertures was 
performed on the long-exposure $R$ image up to an equivalent 
radius of $r\sim$ 11.4 arcsec ($\mu_{T1}= 27.6$ mag arcsec$^{-2}$). 
The centre of the ellipses, the isophotal ellipticity
and position angle were allowed to vary freely, 
until the low surface brightness of the outermost regions
prevented a good convergence of the fits.

The ellipticity is consistent with circular isophotes
out to $r\sim 5$ arcsec, and then it begins to 
increase smoothly until it reaches a maximum value of $\epsilon=0.16$
in the outer regions. The position angle shows a
smooth decrease of $60\degr$ from the innermost 
isophotes (out of the seeing influence region) to the 
outermost ones.

There is no perceptible gradient in the $(C-T1)$ colour profile, 
except that seen at $r<2$ arcsec, likely due to seeing effects.

The brightness profile was successfully fitted by a single S\'ersic
profile in the range $r \sim 2 - 10$ arcsec. This galaxy does not
follow a de Vaucouleurs law, but a shallower one.
 However, we should
stress that we do not have a radial velocity for this object. If it 
was a background galaxy, we would likely lose the central steepening 
of the profile due to seeing and distance effects. 

The effective radius obtained from the analytical profile is in good 
agreement with that calculated from the observed one.

\subsubsection{FS90\,208 (confirmed Antlia member)}

The short-exposure $R$ image was used to obtain an inner profile
up to $r=2$ arcsec. The rest of the profile was obtained from the
long-exposure frame, up to $r=30.1$ arcsec ($\mu_{T1}= 27.5$ mag 
arcsec$^{-2}$). Once more, all the elliptical parameters were 
allowed to vary freely, until a good convergence was prevented due to 
the low surface brightness of the outer isophotes.

This galaxy presents a remarkable change both in ellipticity and in 
position angle at $r= 5.6$ arcsec. The ellipticity decreases from
$\epsilon=0.14$ at $r=2$ arcsec, to $\epsilon=0.02$ at $r=5.6$ arcsec,
and then increases to $\epsilon=0.37$ at $r=23.2$ arcsec. From this point 
outwards, the ellipticity keeps constant. The position angle follows the 
changes in ellipticity, as it varies from $\theta =-22\degr$ at $r=5$ arcsec, 
to $\theta =41\degr$ at $r=7$ arcsec, and then it keeps more or less constant.
These variations are quite similar to those displayed by FS90\,110, albeit
stronger. The B4 coefficient for this galaxy is consistent with its isophotes
being disky.

A small colour gradient is present in the $(C-T_1)$ profile.
The colour gets bluer outwards, from $(C-T_1)=1.8$ at $r=2$ arcsec, 
to $(C-T_1)=1.6$ at $r=10$ arcsec. From this point to the outskirts of the 
galaxy, the colour keeps constant.

As in the case of FS90\,110, the brightness profile of FS90\,208
is better fitted by two components for the same reasons. 
Therefore, we fitted a S\'ersic profile to the 
outer part of the galaxy, in the range $r=12.4 - 24.9$ arcsec, which
was then subtracted in intensities from the original one. We then recovered
the inner component and fitted an independent S\'ersic law
to it (see Fig.\,\ref{ajustes_Sersic}). 

From the integrated $T_1$ 
magnitudes of both components, we found that the inner
one seems to be not so prominent in comparison with the outer, 
as the former is $\sim$ 1.6 times brighter than the latter. 
Contrary to FS90\,110, the radius at which the changes
in ellipticity and position angle arise is smaller than that found 
to possibly separate the two components.

\subsection{Colour maps and unsharp masking}

In order to obtain information about the internal structure of the FS90 cE 
candidates, we built colour maps with their $R$ and $C$ short and long 
exposure images. The individual images were previously filtered by applying a 
median filter with a window size of 5 $\times$ 5 pixels to reduce the noise 
of the maps. 

Following \citet{L06b}, we have also performed an unsharp masking process on 
the long exposure $C$ images of the four galaxies. The $C$ frames were used 
instead of the $R$ ones, as none of the galaxy centres are overexposed in 
the former ones. We have produced circular and elliptical masks,
considering Gaussian kernels with $\sigma$ values in the range 3 -- 30 pixels 
(i.e. 0.8 -- 8 arcsec). To built the elliptical masks, we used the position 
angles and ellipticities of the outer isophotes of each galaxy.

In addition, to test the existence of hidden discs, we have analysed the 
residual images obtained from a fixed ELLIPSE fitting (i.e. by fixing the 
position angle and ellipticity in the whole fitting process).

For FS90\,110 and FS90\,192 we have found no evidence of internal structure.
Both colour maps are smooth and show redder centres than the outskirts, in
agreement to what is seen in their colour profiles (see Fig.\,\ref{Imagenes}). 
As there are FORS1 and ACS images available for them, we will go on  
with this analysis in Sect.\,\ref{masking_HST}.

\begin{figure*}
\includegraphics[width=86mm]{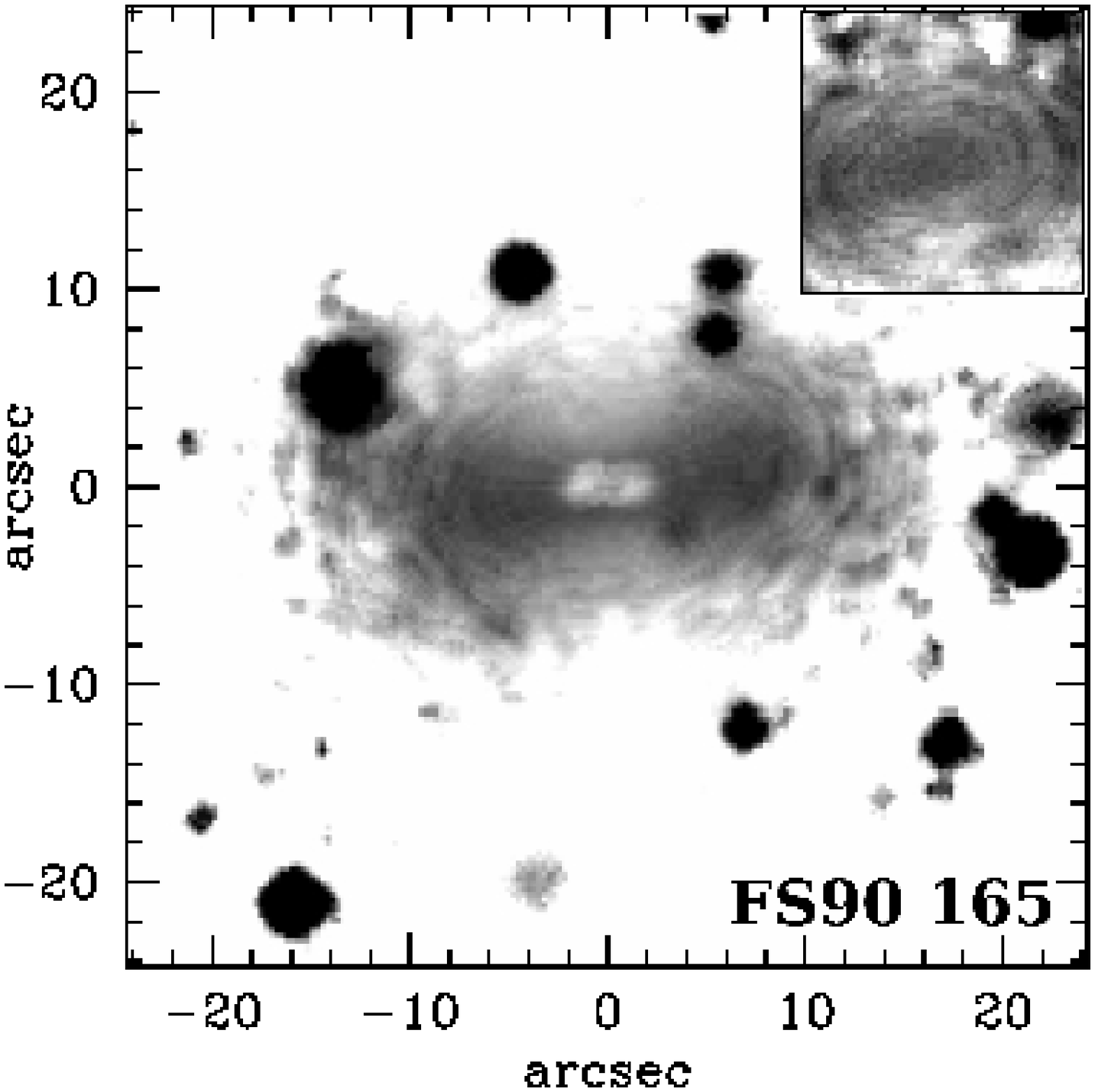}
\includegraphics[width=86mm]{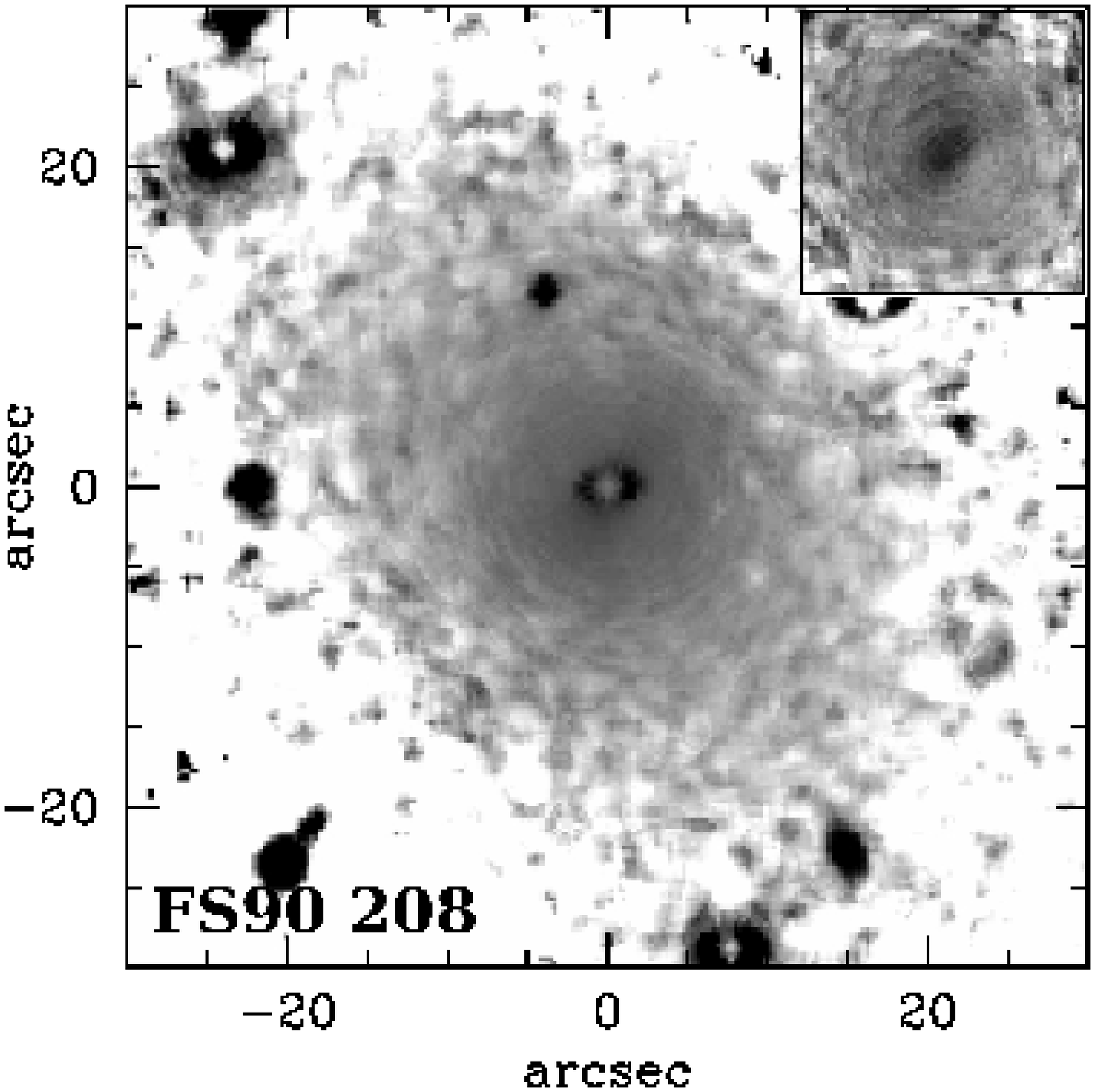}
\includegraphics[width=86mm]{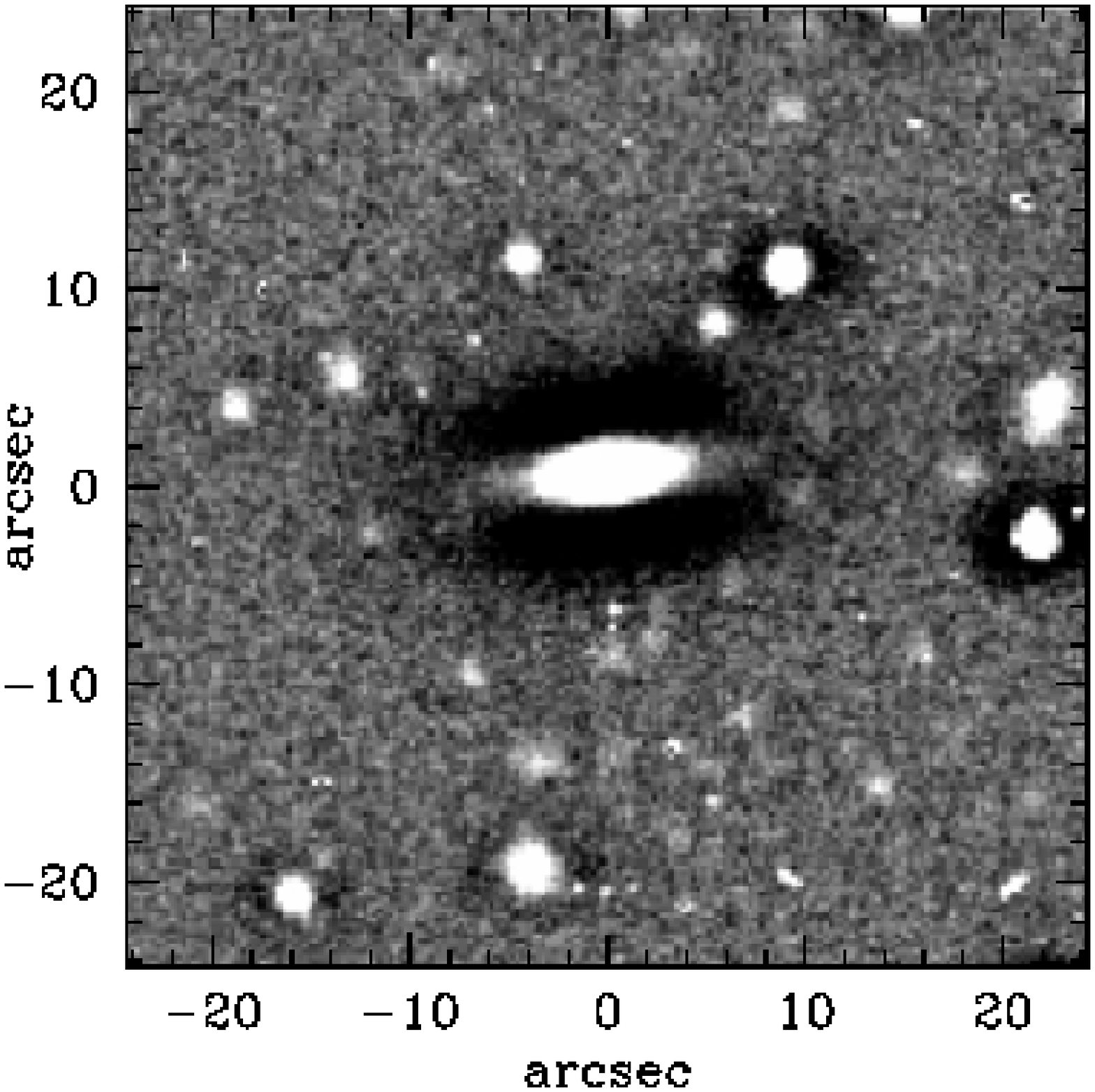}
\includegraphics[width=86mm]{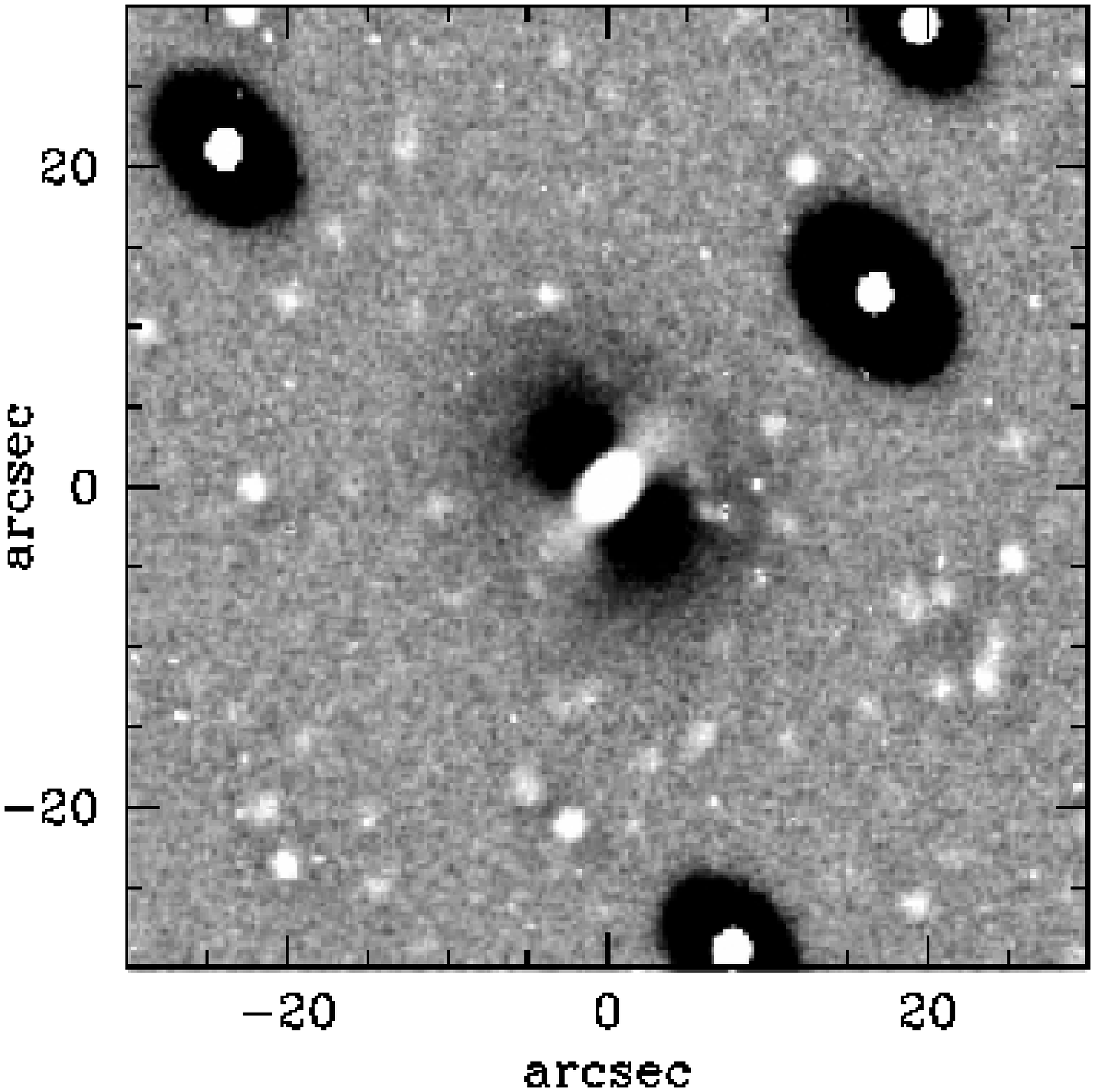}
\caption{{\it Top:} $(C-T_1)$ colour maps of FS90\,165 and FS90\,208.
The large maps were obtained from long exposure images, and the small ones, 
in the upper-right corners, from short exposure frames to recover the 
overexposed central region. The linear scale of the small colour maps is the 
same as in the large ones. A median filter with a window size of $5\times5$ 
pixels was applied to all individual images, previous to the construction of
the maps. The grey scale corresponds to a colour range 
$(C-T_1)=0.0-3.0$ mag, in which black refers to red colours, and 
white to blue ones. {\it Bottom:} Elliptical unsharp masks with kernel size
$\sigma=$ 7 pixels of FS90\,165 ({\it left}) and FS90\,208 ({\it right}).}
\label{Mapas_Mosaico}
\end{figure*}

In Fig.\,\ref{Mapas_Mosaico} we show for FS90\,165 and FS90\,208,  
$(C-T_1)$ colour maps and resulting images of the unsharp masking 
process. It can be seen that the mask of FS90\,165 shows a red disc-like 
structure, not evidenced in its colour profile. This structure is in agreement 
with the existence of disky isophotes, as expected from the positive values
of the B4 index (see Fig.\,\ref{Imagenes}). FS90\,165 seems to be a  
low luminosity lenticular galaxy, displaying a similar colour-map to 
those of similar objects \citep[see, for example,][]{Ch08}. 

The colour map of FS90\,208 is smooth with a small red gradient 
towards the centre, also evident in its colour profile. On the contrary, the 
unsharp mask performed using an elliptical kernel with the same ellipticity
and position angle of its outer component shows a possible bar-like structure, 
that seems to be present in the residual image of a fixed ELLIPSE fitting 
(not shown).

\subsection{Colour-magnitude and luminosity versus mean effective surface 
brightness relations}
\label{K-corr}

\begin{table*}
\caption{Data adopted for the confirmed cE galaxies plotted in Fig.\,\ref{Mueff_Reff}.}
\begin{center}
\begin{tabular}{@{}lcccccccccc}
\hline
{Galaxy} & {$E(B-V)$} & {$(m-M)$} & {$Dist$} & {$R$} & {$r_{\rm eff}$} & {$\langle\mu_{\rm eff}\rangle$} & {$r_{\rm tot}$} & {Companion} & {Dist. Comp.} & {References}\\ 
{}       & {(mag)}         & {(mag)} & {(Mpc)} & {(mag)} & {(arcsec)} &  {\scriptsize(mag arcsec$^{-2}$)} & {(kpc)} & {} & {(kpc)} & {} \\ 
\hline			
M32        & 0.062  & 24.53 & 0.8  & 7.68  & 28.5 & 16.9 & 1.2 & M31       & 5.5 & 1,2,3\\
NGC\,4486B & 0.021  & 31.26 & 17.9 & 12.72 & 2.5  & 16.7 & 1.0 & NGC\,4486 & 28  & 4,5,6\\
NGC\,5486A & 0.055  & 32.08 & 26.1 & 13.07 & 4.1  & 18.1 & 1.6 & NGC\,5486 & 3.1 & 6,7,8\\   
A496cE     & 0.138  & 35.70 & 138.0& 17.35 & 0.7  & 18.6 & -   & A496 cD   & 14  & 9\\
\hline
\end{tabular}
\end{center}
\medskip

{\it Notes.-} Extinction values are from \citet{S98}. References:  
(1) \citet{M98} and references therein; (2) \citet{G02}; (3) \citet{choi02}; 
(4) \citet{A03}; (5) \citet{ton01}; (6) \citet{N87}; (7) \citet{Ma05}; (8) 
\citet{dV91}; and (9) \citet{Ch08}. In all cases, the mean effective 
surface brightness ($\langle\mu_{\rm eff}\rangle$) was obtained from 
the effective radius ($r_{\rm eff}$), and $R$ magnitudes transformed into 
$T_1$ ones through $R-T1=-0.02$ (see Section\,\ref{observations}), 
with eq.\,1 of Paper I. For A496cE, the $R$ magnitude was 
calculated from its $r'$ one \citep{Ch08}, through $(r'-R_c)=0.25$ 
\citep{F95}.
\label{M32}
\end{table*}

\begin{figure*}
\includegraphics[width=84mm]{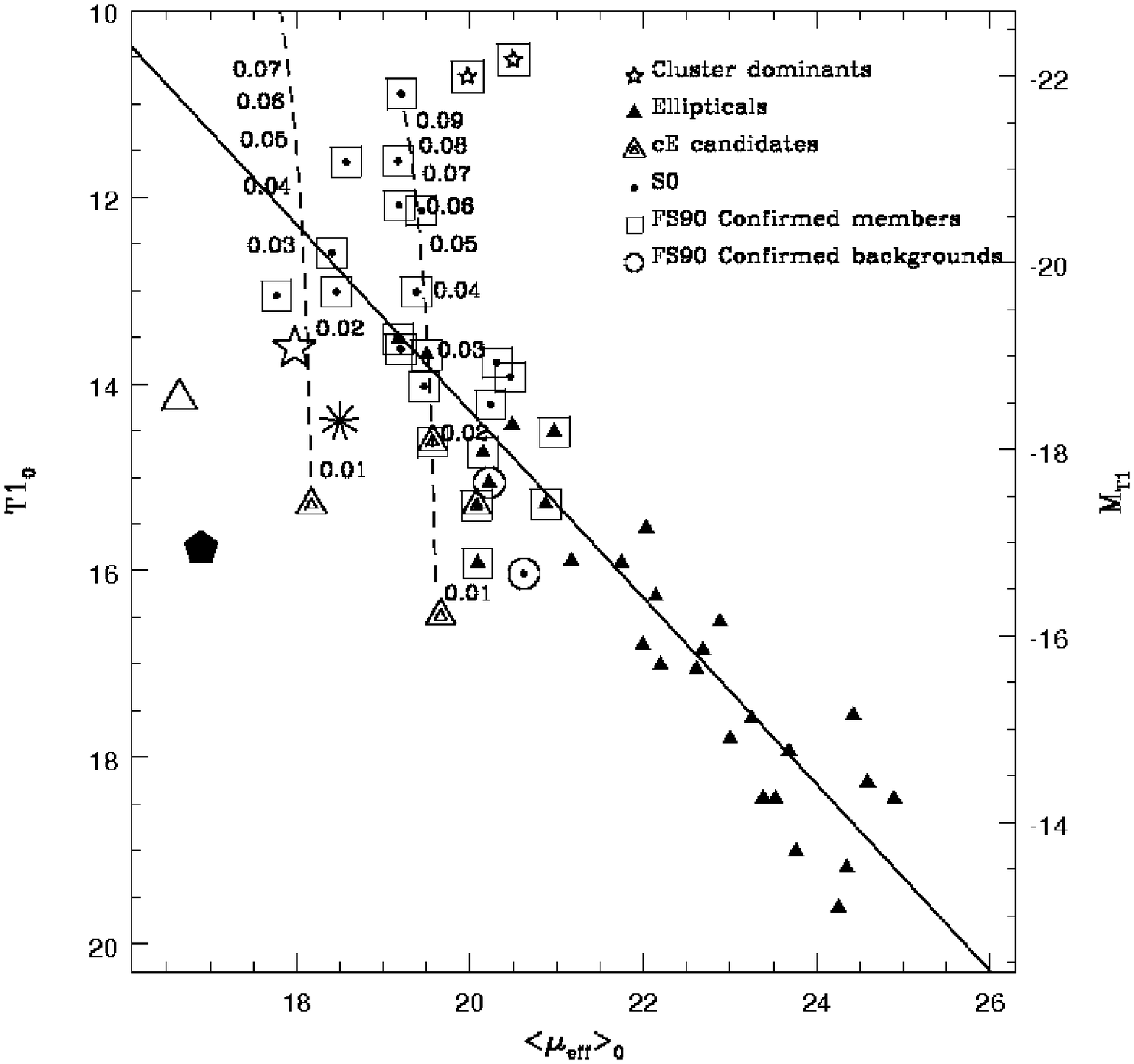}
\includegraphics[width=84mm]{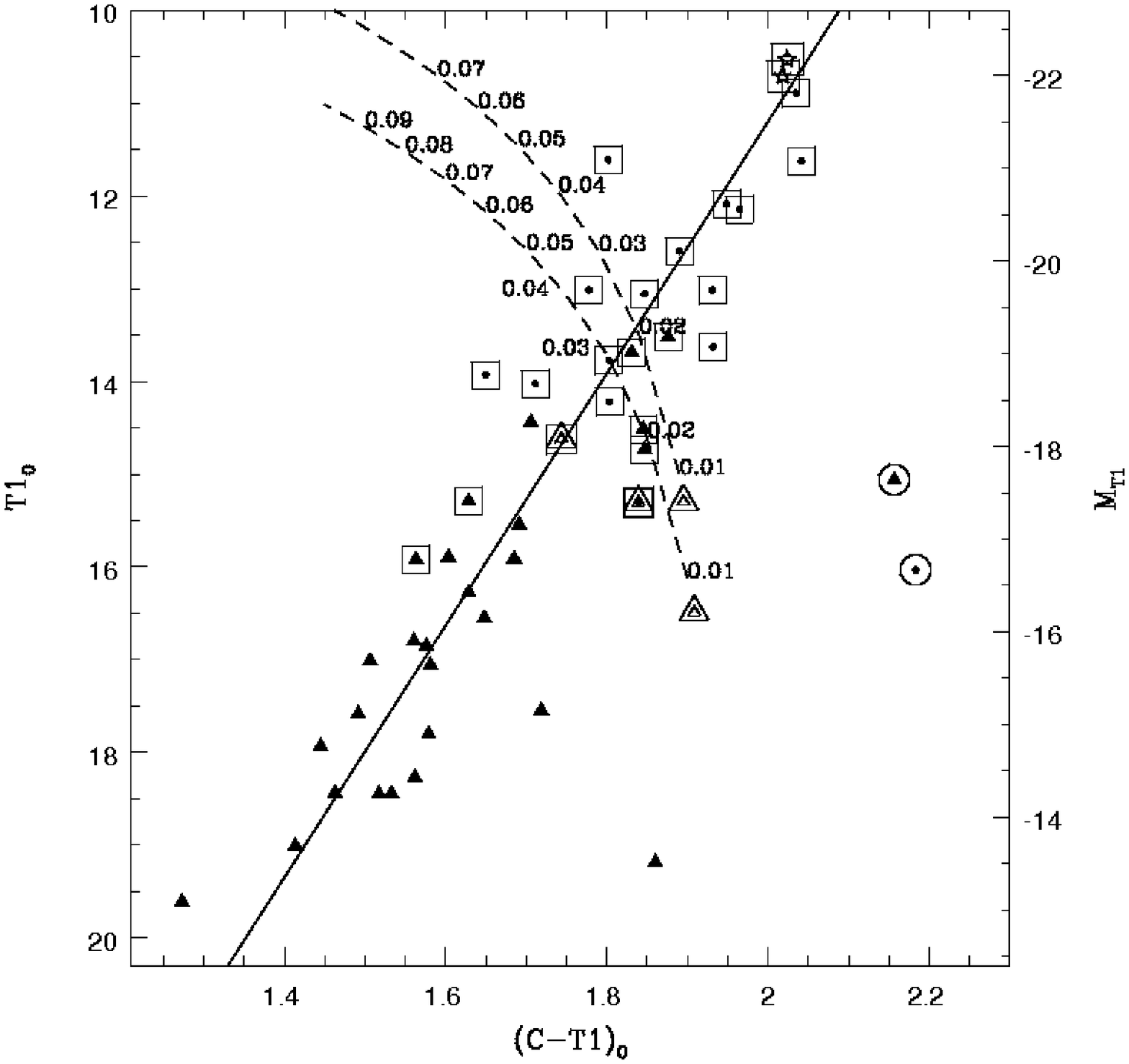}
\caption{{\it Left:} $T_1$ magnitude vs. mean effective surface brightness 
  plot where different morphologies are indicated. The solid line is the
  locus of constant effective radius ($\sim$ 1 kpc) of
  Antlia early-type galaxies with $T1>13$ mag (see Paper I). Two
  FS90 cE candidates set the lowest limit in effective radii of our 
  Paper I sample. All galaxy positions in the plot correspond to their 
  observed apparent magnitudes, except for M32 ({\it big filled 
  pentagon}), NGC\,4486B ({\it big open triangle}), NGC\,5846A 
  ({\it big star}), and A496cE ({\it big asterisk}), which show the 
  location they would display if they were at the Antlia distance. 
  The dashed lines show the possible locations that FS90\,110 and 
  FS90\,192 would hold at the Antlia distance, if their true distances 
  corresponded to the redshift range $z=0.01 - 0.1$ (small numbers close to 
  the dashed lines indicate such redshifts). {\it Right:} colour - magnitude 
  diagram including the FS90 cE candidates (symbols like in the left panel). 
  The solid line shows the mean colour-magnitude relation followed by 
  early-type Antlia galaxies (Paper I). The dashed lines correspond to the 
  ones displayed in the left panel. The absolute magnitude scale (on the right 
  of both panels) is only valid for Antlia confirmed members.}
\label{Mueff_Reff}
\end{figure*}

The left panel of Fig.\,\ref{Mueff_Reff} shows a $T_1$ vs. 
$\langle \mu_\mathrm{eff}\rangle$ (absorption corrected) plot for the FS90 
early-type galaxies from our Paper I sample, where their FS90 morphology has 
been indicated. We also depict, as a reference, the locus of constant effective 
radius followed by galaxies fainter than $T_1=13$ mag (see Paper I). 
In this plot, the locations of all these galaxies correspond to observed 
apparent magnitudes. For comparison, we have also included some confirmed cE
galaxies that display the positions they would have at the Antlia distance. 
These are M32 ({\it big filled pentagon}), NGC\,4486B 
({\it big open triangle}), NGC\,5846A ({\it big star}), 
and A496cE ({\it big asterisk}) (see Table\,\ref{M32}).

In this figure, 
FS90\,165 and FS90\,208 share the locus of the Antlia early-type 
galaxies. On the contrary, FS90\,110 and FS90\,192 are located far from 
this relation defined by early-type galaxies, towards higher mean 
effective surface brightness, smaller effective radius or fainter magnitudes. 

In particular, FS90\,110 presents the highest mean effective surface 
brightness of the whole Paper I sample, with the exception of FS90\,94, 
a bright S0 Antlia member. Furthermore, FS90\,110 seems to 
extend the sequence followed by bright elliptical and S0 galaxies 
(in a perpendicular direction with respect to the early-type galaxies' 
relation), in the same way as M32 would do if it were placed at the Antlia 
distance. This latter correlation corresponds to the \citet{K77} 
scaling relation followed by bright ellipticals and bulges of spiral 
galaxies on the $r_{\rm eff}$ vs. $\mu_{\rm eff}$ plot, which is a projection 
of the Fundamental Plane \citep{DD87}. NGC\,4486B, NGC\,5846A 
and A496cE are located within this sequence as well, towards smaller 
effective radius. However, M32 is the most extreme case. This 
behaviour of cE galaxies has already been noticed by, for instance, 
\citet{N87} \citep[see also][]{Ch07}. FS90\,192 does not follow the same 
trend as the cEs do.

The right panel of Fig.\ref{Mueff_Reff} shows the colour-magnitude diagram 
of the FS90 early-type galaxies from the Paper I sample, where the 
FS90 cE candidates have been included. The solid line 
shows the mean colour-magnitude relation followed by early-type Antlia
galaxies. It can be seen that FS90\,110 is located on the red border of the 
relation, at almost the same position as FS90\,165, and 
evidently separated from the early-type background galaxies with similar 
apparent magnitudes. On the contrary, FS90\,192 clearly deviates from the 
colour-magnitude relation of Antlia members towards redder colours, or fainter 
magnitudes. 

We note that, considering the $(B-R)$ vs $M_B$ 
colour-magnitude relation of the Perseus cluster \citep*{Con02}, and adopting 
for M32 $(B-R)\sim 1.45$ mag \citep{P93} and $M_R\sim -16.8$ (see 
Table\,\ref{M32}), the location of M32 in this diagram presents the same trend 
like FS90\,110 and FS90\,192 in ours, i.e. shifted towards redder colours or 
fainter magnitudes from the cluster mean colour-magnitude relation. 

As an additional test, we assume that FS90\,110 and FS90\,192 are elliptical 
background galaxies that `fall off' the mean relations followed by Antlia 
early-type members, due to the effect of distance. Then we can calculate  
the shift of their positions in the $\langle \mu_{\rm eff} \rangle$ vs. 
$r_{\rm eff}$ and colour - magnitude plots applying cosmological dimming 
and K-corrections to their observed surface brightnesses, and luminosities and 
colours, respectively. From \citet{F95} we adopt the K-correction to $(B-R)$
for an elliptical galaxy. Assuming that early-type galaxies are old stellar
systems, we transformed this correction into $(C-T1)$ following \citet{FF01}. 
The dashed lines in both panels of Fig.\,\ref{Mueff_Reff} show the different 
positions that these galaxies would hold at the Antlia distance, if their real 
distances were within a redshift range $z=0.01 - 0.1$. Please note that a 
redshift of $z \approx 0.01$ corresponds approximately to the adopted Antlia 
distance.

Fig.\,\ref{Mueff_Reff} shows that, to fall on the locus of the Antlia 
relation, FS90\,110 might have, at most, an intrinsic luminosity similar to 
those of bright early-type Antlia members, then being at a distance of 
$\sim$ 120 Mpc ($z \approx 0.03$). This shift would place it within 
the colour-magnitude relation at the right panel. 

For FS90\,192 two options 
exist in the $\langle \mu_{\rm eff} \rangle$ vs. $r_{\rm eff}$ plot: it could 
be a bright dwarf elliptical at $\sim$ 120 Mpc, or a giant elliptical 
galaxy placed as far as $\sim$ 400 Mpc ($z \approx 0.1$). However, from the 
colour - magnitude diagram we can see that the second possibility must be 
discarded as, for that redshift, FS90\,192 would be completely out of the 
mean relation. 

Regarding FS90\,192, the images do not show any concentration of 
galaxies in its neighbourhood, which would make it an interesting object even 
if it were not confirmed as an Antlia cE galaxy: isolated ellipticals of 
moderate luminosity are not common objects. In the case of FS90\,110, we can 
not rule out from the present analysis that it may be an ordinary background 
galaxy.

\section{Analysis of  FORS1 and ACS data}
\label{HST}

In order to obtain more evidence in favour of, or against   
FS90\,110 and FS90\,192 being cE galaxies, we have analysed two
frames obtained with FORS1, centred on NGC\,3258 and NGC\,3268, 
respectively. As said above, these frames are deeper and were taken 
under better seeing conditions than the MOSAIC ones. We have also
used two ACS images from the HST archive also centred on NGC\,3258 and 
NGC\,3268, respectively. Unfortunately, there are no similar data available 
for FS90\,165 and FS90\,208.

\subsection{Low surface brightness bridge}

\begin{figure}
\includegraphics[scale=0.44]{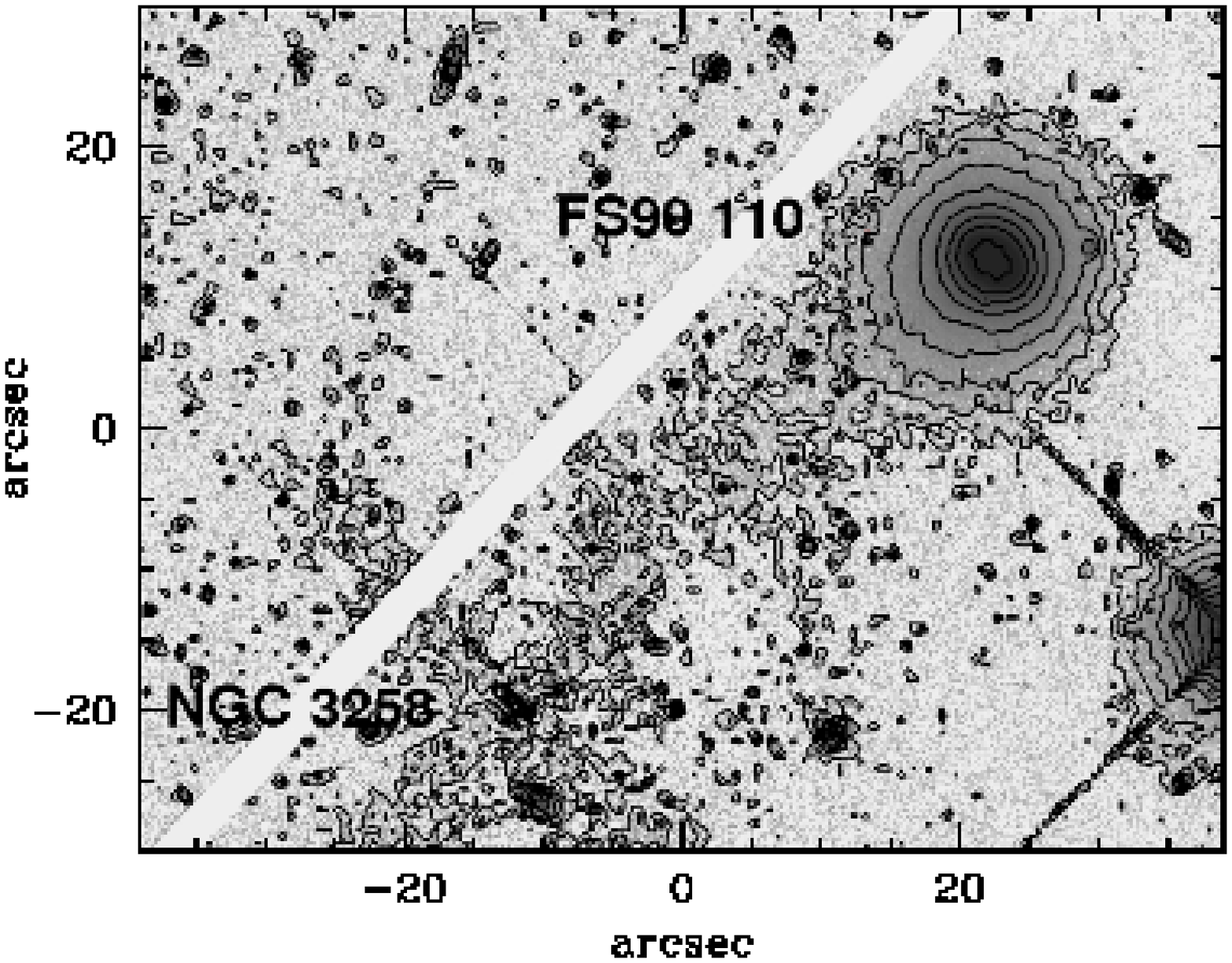}
\includegraphics[scale=0.44]{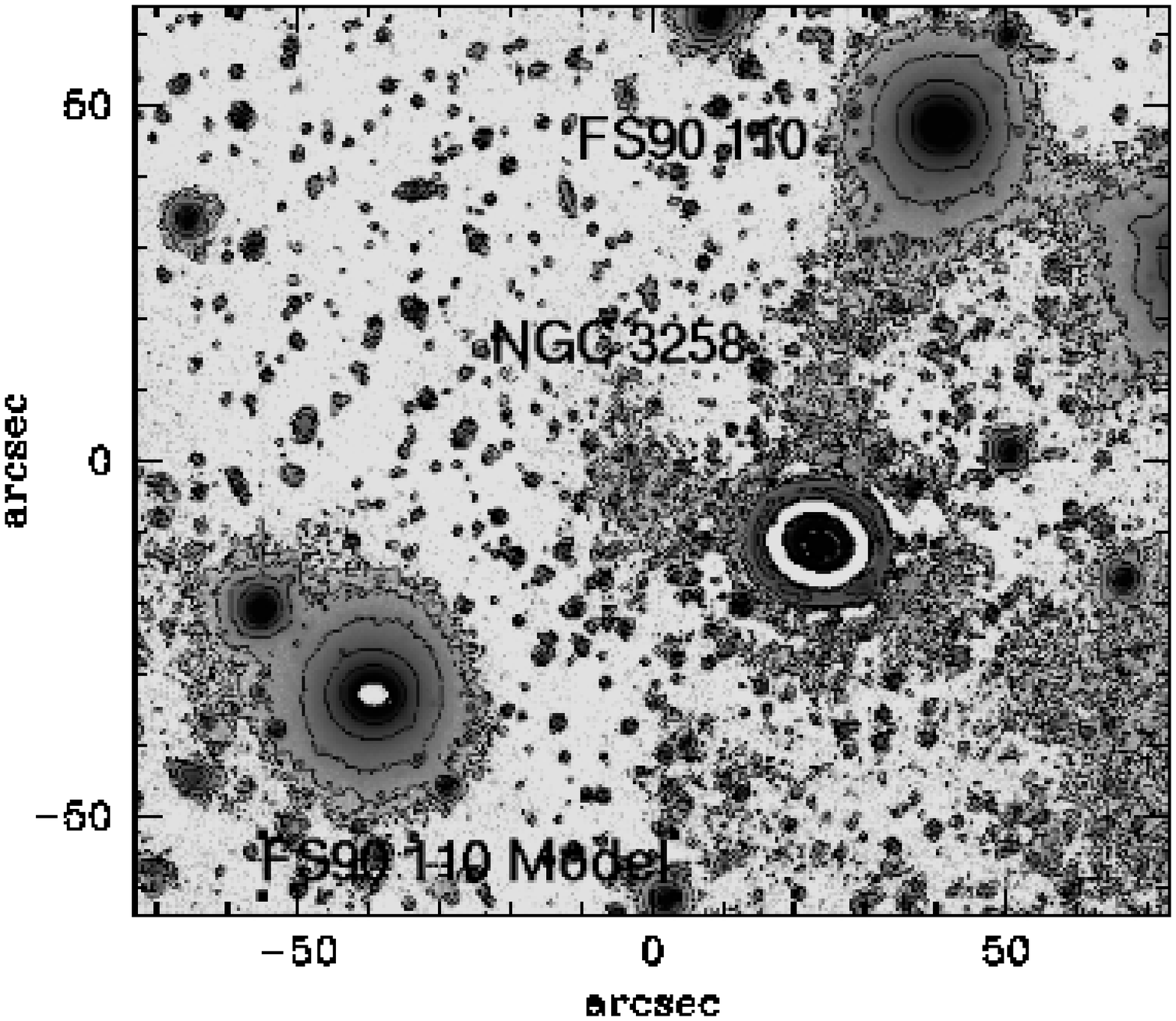}
\caption{Contour maps showing the low surface brightness structure that 
seems to link FS90\,110 to NGC\,3258. {\it Top:} F814W ACS frame from 
which a model of NGC\,3258 has been subtracted. {\it Bottom:} To test if 
the low surface brightness `bridge' is an artifact 
due to the subtraction of the light of NGC\,3258, we have added a model of 
FS90\,110 to the $V$ FORS1 image, shifted in position angle 
from its original location but keeping its galactocentric distance to the 
centre of the dominant galaxy. After the
subtraction of the NGC\,3258 light model, there is no similar low surface 
brightness structure left around the artificial FS90\,110 galaxy.}
\label{Detalles2}
\end{figure}

In the FORS1 and ACS images of FS90\,110 and FS90\,192 we have 
not detected outer structures different from those previously seen on the  
MOSAIC images. The very low surface brightness `bridge' that seems to link 
FS90\,110 to NGC\,3258 is detected in both the $V$ and $I$ FORS1 images 
as well as  in the F814W ACS frame. No comparable structures were found in 
the corresponding images of FS90\,192. Its outer isophotes look quite regular 
without  visible evidence of distortion.

In the top panel of Fig.\,\ref{Detalles2} we show brightness contour levels 
superimposed on the F814W ACS frame of FS90\,110 from which a model of
NGC\,3258 has been subtracted. In order to test if the low surface 
brightness structure could be just an artifact due to the  subtraction of 
NGC\,3258, we added to the original $V$ FORS1 image of NGC\,3258, a model 
of FS90\,110 obtained with BMODEL. This artificial galaxy with similar 
characteristics as FS90\,110, was placed in a different position but keeping 
the original galactocentric distance to the centre of the dominant galaxy. We 
subtracted the light of NGC\,3258 from this new image, and no low surface 
brightness `bridge' was found at the location of the added object 
(see Fig.\,\ref{Detalles2}). This result can be understood as 
a proof that such `bridge' is not a spurious detection due to image 
processing. Another argument supporting this statement is the fact that
this low surface brightness structure does not connect the two galaxy
centres, as can be seen in both panels of Fig.\ref{Detalles2}.

\subsection{Colour maps and unsharp masking}
\label{masking_HST}

With the purpose of getting additional information on the possible internal 
structure of FS90\,110 and FS90\,192, we built $(V-I)$ and $(B-I)$ 
colour maps from the FORS1 and ACS images, respectively. We have also 
performed unsharp masking on those frames with the best signal-to-noise 
ratio, i.e. the FORS1 $V$ and ACS $I$ images. Elliptical masks were 
constructed in a similar way as those of MOSAIC.

\begin{figure*}
\includegraphics[width=86mm]{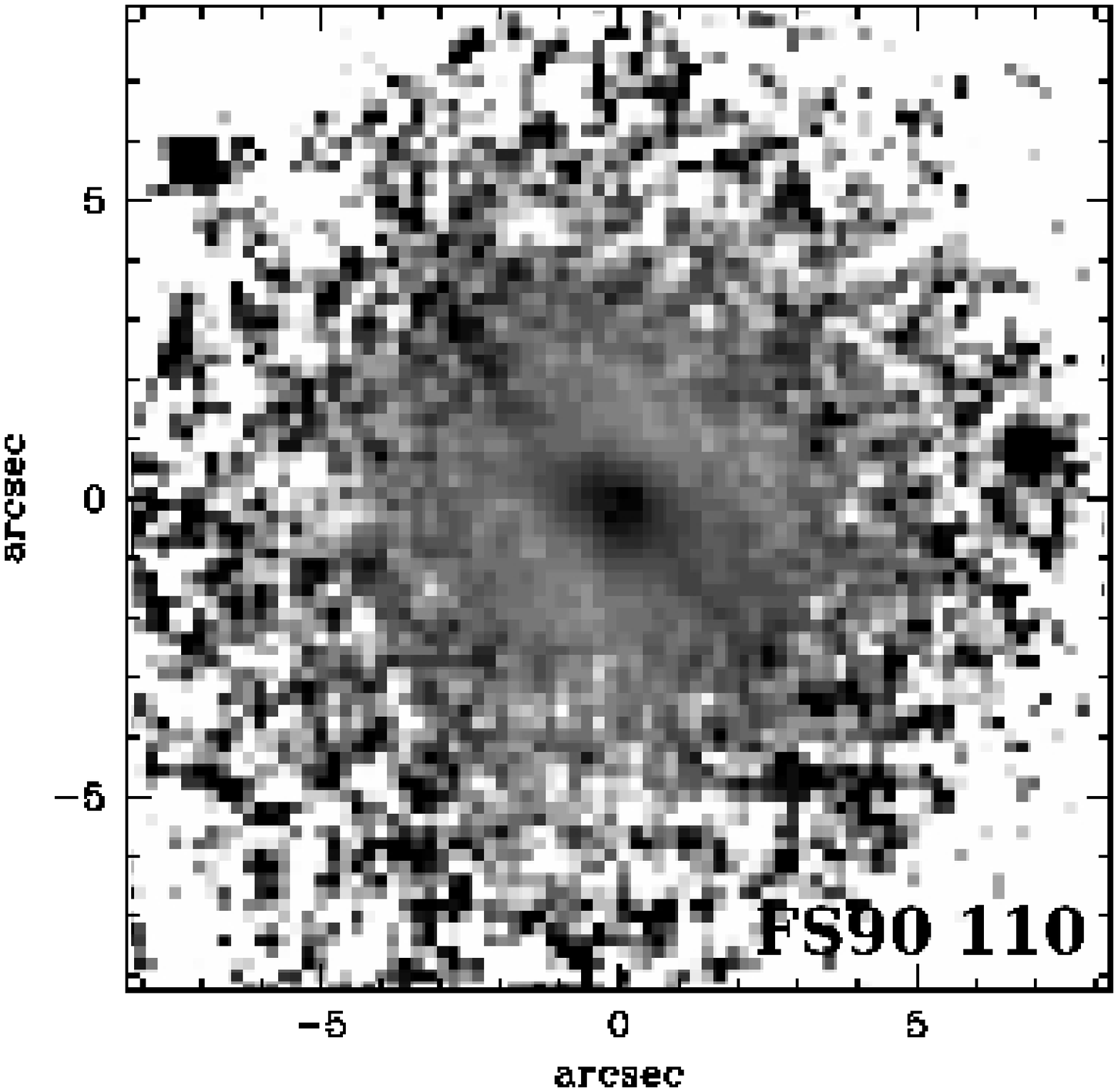}
\includegraphics[width=86mm]{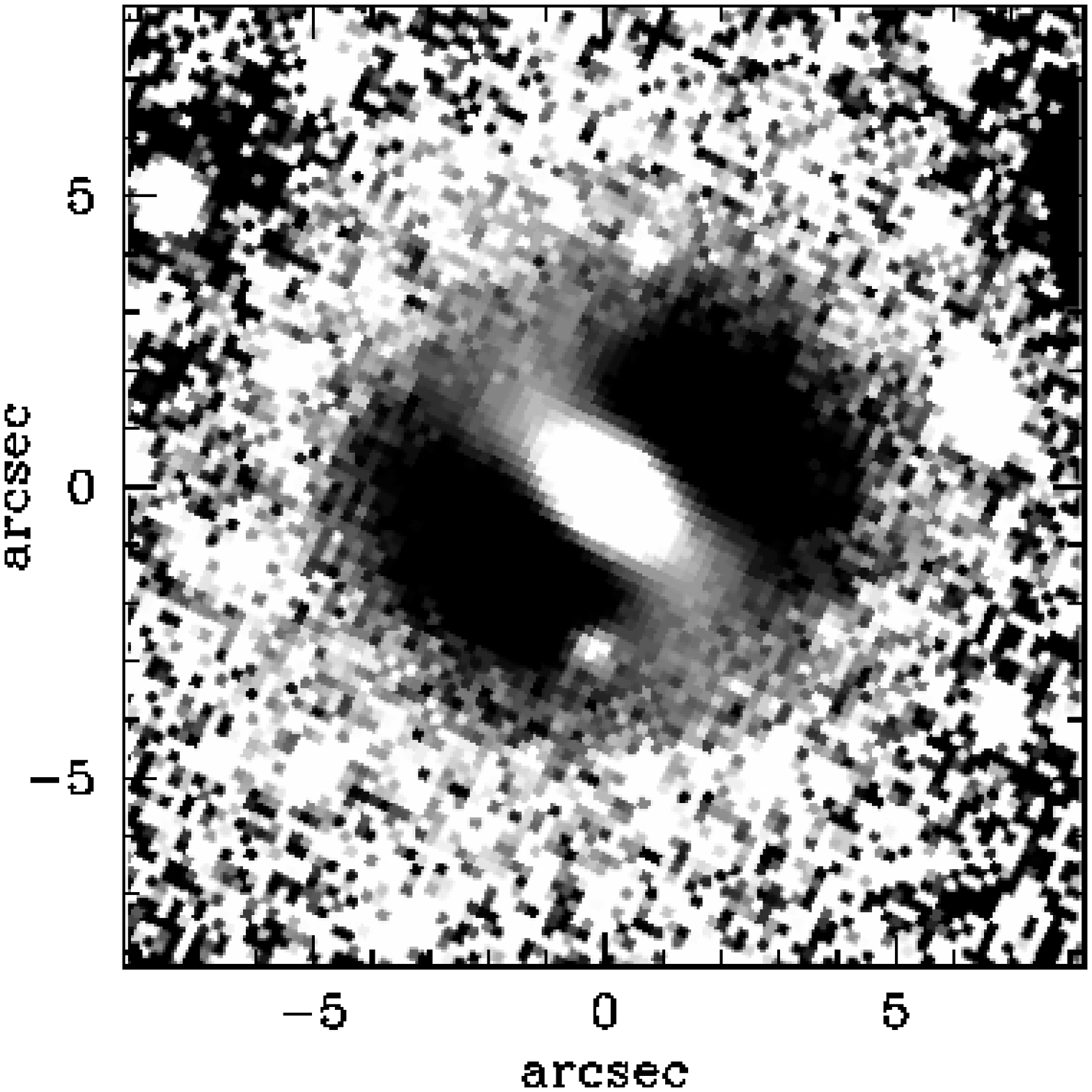}
\caption{{\it Left:} $(B-I)$ colour map of FS90\,110,  
obtained from ACS images. A median filter with a window size of 
$5\times5$ pixels was applied. The displayed grey scale corresponds
to a colour range $(B-I)=0.0-2.5$ mag, in which black refers 
to red colours, and white to blue ones. {\it Right:} FORS1 elliptical
unsharp mask with a Gaussian kernel size of $\sigma=$ 5 pixels 
(i.e. 1 arcsec).}
\label{Mapas_HST}
\end{figure*}

In Fig.\,\ref{Mapas_HST} we show the ACS $(B-I)$ colour map of FS90\,110 
(left panel), which has a better spatial resolution than the FORS1 one. 
A FORS1 unsharp elliptical mask is also displayed (right panel), resulting
from a $\sigma$ value of 5 pixels (i.e. 1 arcsec). 
In the inner region of FS90\,110, we  see an embedded warped structure 
which is detected on both the colour map and the mask. This structure is 
redder than the rest of the galaxy.

Although the ACS images show a higher spatial resolution, 
we prefer to show a FORS1 mask, as the former contains an artifact 
in the same location where the inner structure is expected to show up. All
bright point sources also show this artifact (note the bright star at the 
right side of the top panel of Fig.\,\ref{Detalles2}.)

In the colour maps and unsharp masks of FS90\,192 (not shown) we have not 
found any structure different from  what has been already seen in the 
MOSAIC data.

\subsection{Brightness profiles}

\begin{table}
\caption{FS90 cE candidates located in the FORS1 images.} 
\setlength\tabcolsep{1.5mm}
\label{tabla_VI}
\begin{tabular}{@{}ccccr@{}c@{}lcr@{}c@{}l}
\hline
\multicolumn{1}{c}{FS90}  &  \multicolumn{1}{c} {$V$} &  \multicolumn{1}{c}{$(V-I)$}  & \multicolumn{1}{c} {$\mu_{_{V}}$} &   \multicolumn{3}{c} {$r_{_{V}}$} &  \multicolumn{1}{c} {$\langle \mu_{\rm eff}\rangle_{_{V}}$} &  \multicolumn{3}{c} {$r_{\rm eff_{_{V}}}$} \\
\multicolumn{1}{c}{ID}  & \multicolumn{1}{c}{\scriptsize mag}& \multicolumn{1}{c}{\scriptsize mag} & \multicolumn{1}{c}{\scriptsize mag arcsec$^{-2}$} & \multicolumn{3}{c}{\scriptsize arcsec} & \multicolumn{1}{c}{\scriptsize mag arcsec$^{-2}$} & \multicolumn{3}{c}{\scriptsize arcsec} \\
\hline
110  & 16.41  & 1.14  & 27.7 & 12&.&3 & 18.8 & 1&.&2  \\
192  & 18.00  & 1.21  & 27.3 &  8&.&6 & 21.0 & 1&.&6  \\
\hline
\end{tabular}
\medskip

{\it Notes.-} $\mu_{_V}$ gives the surface brightness of the outermost 
isophote within which integrated magnitudes and colours are measured. 
$r_{_{V}}$ is the equivalent radius ($r=\sqrt{a\cdot b}$) of that isophote. 
$\langle \mu_{\rm eff}\rangle$ is obtained from $r_{\rm eff}$. 
All these values were obtained from ELLIPSE. 
\end{table}

\begin{figure*}
\center
\includegraphics[width=80mm]{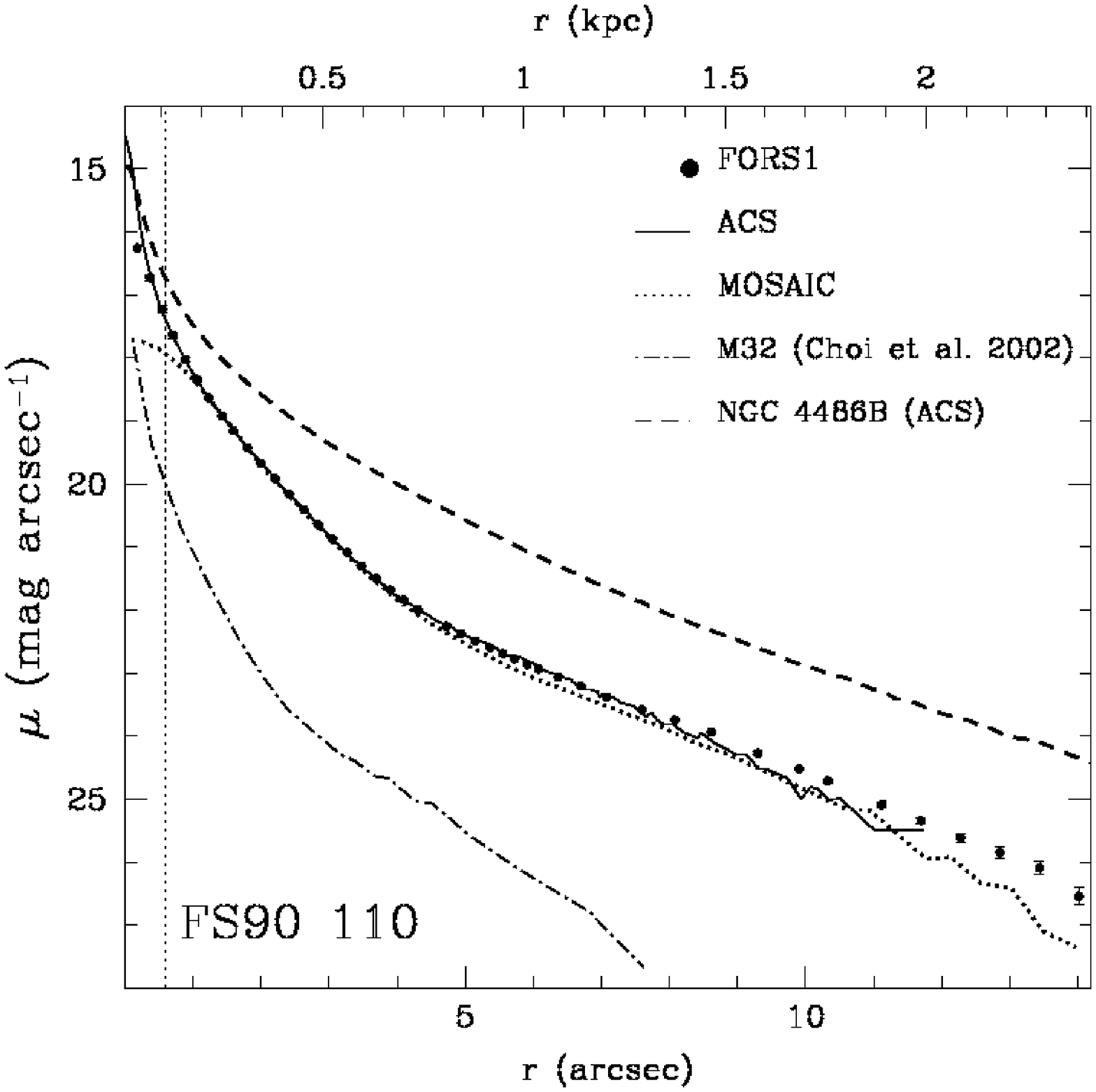}
\includegraphics[width=80mm]{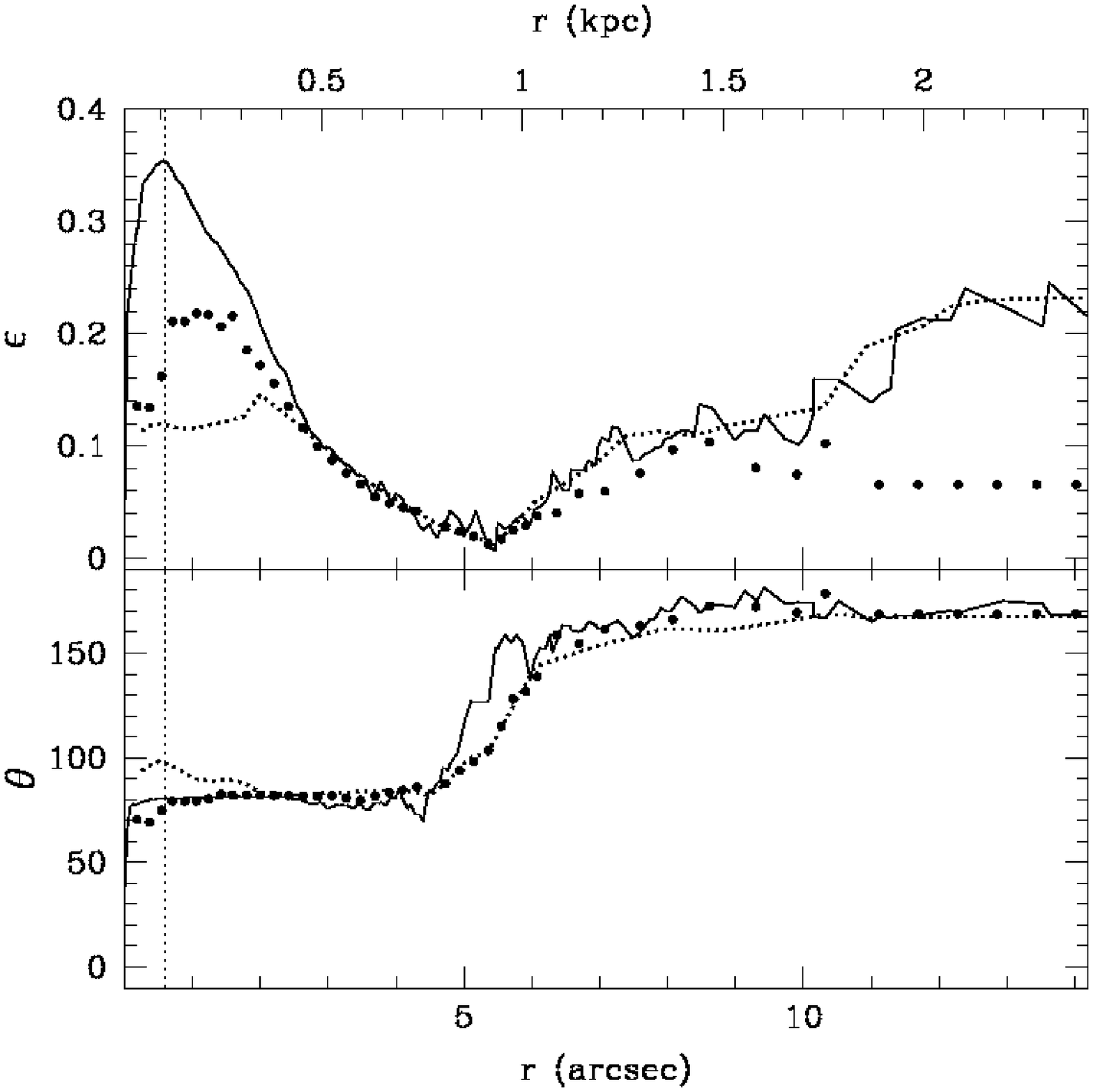}\\
\includegraphics[width=80mm]{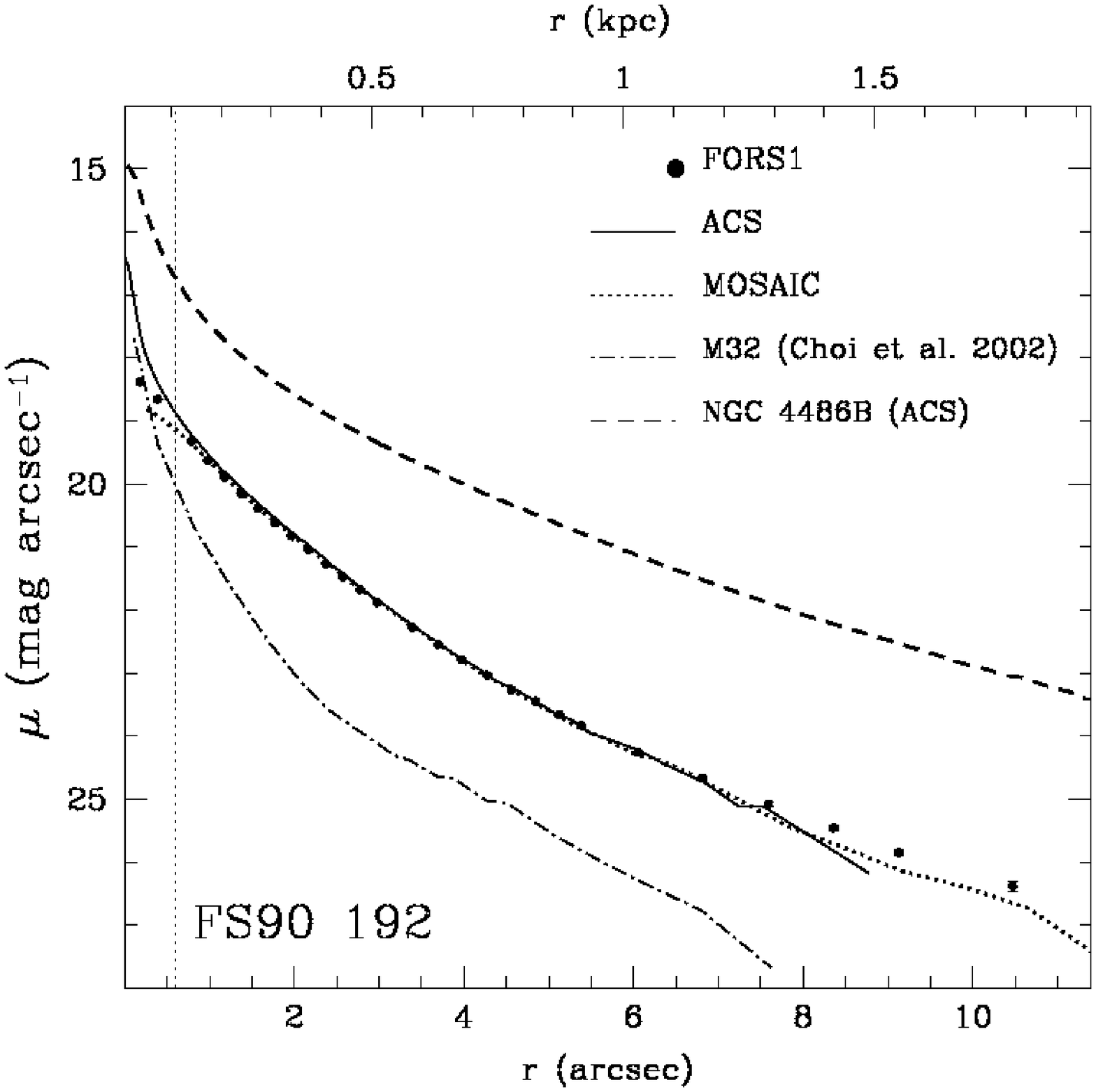}
\includegraphics[width=80mm]{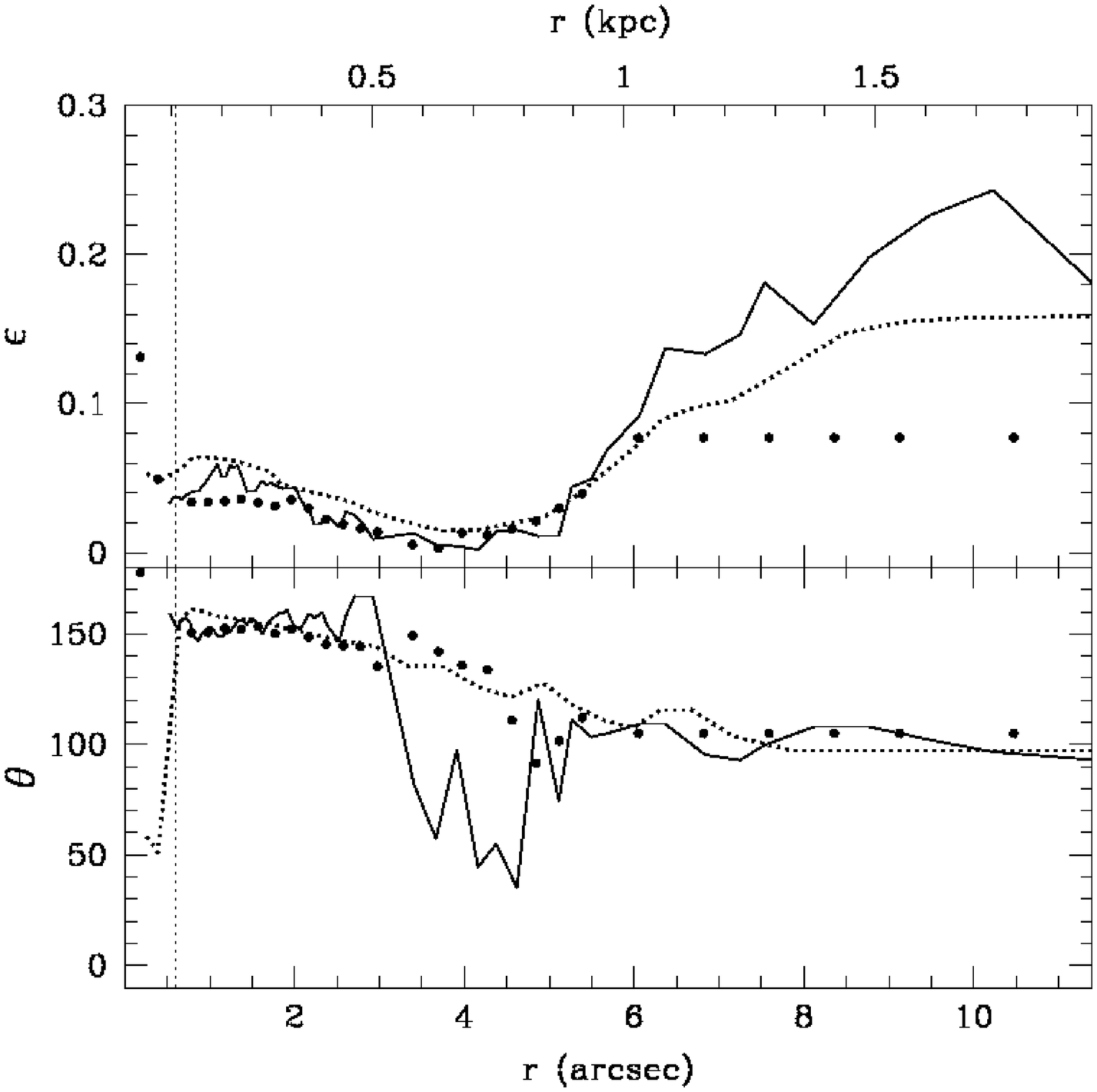}
\caption{{\it Left:} $T_1$ brightness profiles of FS90\,110 (top) and 
FS90\,192 (bottom), compared with the $V$ one obtained from FORS1, 
and the $I$ one obtained from ACS. The FORS1 profiles have been 
shifted by adding 1.1 mag in order to match the $T_1$ profiles, and the ACS 
ones by adding 0.4 mag. We have also added the $z$ ACS brightness profile 
of NGC\,4486, and that of M32 in the $I$ band from \citet{choi02}, with 
their equivalent radius re-scaled to the Antlia distance. For 
clarity, the M32 profile has been shifted by adding 2.5 mag. {\it Right:} 
Comparison of ellipticity and position angle vs. equivalent radius of FS90\,110
and FS90\,192, obtained from MOSAIC, FORS1 and ACS images (symbols like
in the left panel). The vertical dotted lines in all panels show the seeing 
region of the FORS1 images. The equivalent radius scale displayed at the 
top of all panels was obtained considering the adopted Antlia distance (1 
arcsec $\simeq$ 170 pc).}
\label{I_110}
\end{figure*}

Figure.\,\ref{I_110} shows a comparison of the MOSAIC $T_1$ 
brightness profiles of FS90\,110 and FS90\,192, with the $V$ and 
$I$ ones obtained from FORS1 and ACS, respectively.  
The photometric parameters obtained from the VLT $V$ and $I$ 
profiles are listed in Table\,\ref{tabla_VI}. 

Most of the features observed in the MOSAIC profiles are
confirmed through VLT and HST images, including the ellipticity and 
position angle variations. The exception is the ELLIPSE B4 coefficient 
behaviour in FS90\,110, as it does not become positive at $r\sim6$ arcsec. 
The high central surface brightness displayed by FS90\,110 in the ACS profile 
is noticeable, detected thanks to the low FWHM of these images.

In order to test the values of the $N$ S\'ersic indices obtained for 
the $T_1$ brightness profile (see Sect.\,\ref{profiles}) of FS90\,110, we 
have fitted two coupled general S\'ersic laws to its ACS $I$ brightness 
profile, within the equivalent radius range 2.0 -- 11.8 arcsec. The S\'ersic 
index N (i.e. 1/n) for both components resulting from this new fit 
(1.8$\pm$0.2 and 1.25$\pm$0.05 for the outer and the inner components, 
respectively), are in quite good agreement with those obtained from the MOSAIC 
profile (see Table\,\ref{ajustes_mosaic}). 

We intended to perform a two component fit over the whole useful range
of the ACS profile, i.e. 0.3 -- 11.8 arcsec, as the innermost 0.3 arcsec are
expected to be affected by the seeing. However, no model could be obtained that 
properly fitted the whole profile considering an inner cut radius 
lower than 1.5 arcsec ($>$ 2 FWHM). This indicates that two S\'ersic 
laws can represent the light profile of the galaxy 
only if the central region is excluded from the fit. Such limitation of 
S\'ersic profiles could not be avoided by, for example, performing a PSF 
convolution with the model.

\citet{choi02} have obtained the $I$ profile of M32, which is shown 
in Fig.\,\ref{I_110} with its equivalent radius re-scaled to the
Antlia distance. In addition, we have obtained the ACS $z$ profile
of NGC\,4486B which is also shown re-scaled.  
The profiles of FS90\,110 and M32 look quite similar both 
in shape and high central surface brightness, albeit the former extends 
over a larger galactocentric radius. 
M32 also displays significant changes both in ellipticity 
and in position angle \citep{choi02}, similar to those found in FS90\,110. 
On the other hand, NGC\,4486B is larger than FS90\,110 and does not show a 
two component profile, similarly to FS90\,192. This is in agreement 
with \citet{F06}, who fitted a single component S\'ersic law to the profile of 
NGC\,4486B.

\section{Discussion and Conclusions}
\label{conclusions}

\subsection{FS90\,165}

From the analysis performed in the previous section, we conclude
that FS90\,165, an spectroscopically confirmed Antlia member,
cannot be considered as a cE galaxy. It follows fundamental relations 
of low surface brightness early-type Antlia galaxies (see also Paper I), 
is larger than any confirmed cE object, and presents no compact
morphology. It is also located far away in projection from any bright galaxy, 
and from the central cluster region, a characteristic shared by all  
confirmed cE galaxies. Its $T_1$ brightness profile is well fitted by 
a single S\'ersic law along its whole radius range, i.e. it does not show two 
components. However, the S0 morphological classification 
assigned to this galaxy by FS90 is in agreement with the B4 coefficient
values that point to disky isophotes, and with the red inner disc
found in its $(C-T_1)$ colour map and confirmed through unsharp masking.
There are no images from VLT or HST for this object. 

\subsection{FS90\,208}

Certainly, FS90\,208 is not a cE galaxy either. It is a confirmed 
Antlia member that follows the same fundamental relations as
FS90\,165 and the rest of low surface brightness early-type galaxies, and
it is not compact. Besides, it is not close in projection to any bright galaxy 
and is far from the centre of the cluster. However, it displays a two component 
brightness profile and presents important variations in ellipticity and 
position angle that resemble those shown by M32 \citep{choi02}, albeit 
stronger. Curiously, these variations arise at the same galactocentric radii 
than those displayed by FS90\,110, and behave in a similar way. FS90 have 
classified this object as an S0 galaxy, which is in agreement with
the two components present in its brightness profile, as well as with the B4 
coefficient which indicates that the isophotes are nearly disky. Moreover, a 
process of unsharp masking reveals what seems to be an inner bar structure, 
slightly detected with the subtraction of a fixed ELLIPSE model of the galaxy 
from the MOSAIC image. However, a triaxial object in which the axial ratios 
vary with radius, could also display isophote twisting \citep{BM98}.

\subsection{FS90\,192}

There is no radial velocity available for FS90\,192. It displays no 
perceptible colour gradients and, if it was an Antlia member, it would 
be as compact as other cE galaxies. Furthermore, its projected distance 
to NGC\,3268 would be smaller than that between NGC\,4486B and NGC\,4486.
Although it does not display two components in its brightness profile, it is
worth noting that NGC\,4486B in the Virgo cluster presents a single component 
profile as well and a similar n (i.e. 1/N) value.

Despite all the interesting features mentioned above, we cannot confirm nor
 rule out that this object is in the background. It does not follow the 
early-type Antlia members fundamental relations, as it appears shifted towards 
higher surface brightnesses, fainter magnitudes, or redder colours. M32 would 
hold a comparable position on the colour-magnitude diagram if it were placed at
the Antlia distance. However, FS90\,192 does not follow the relation between 
mean effective surface brightness and luminosity defined by the confirmed cEs. 
Under the hypothesis that it is a background object, FS90\,192 would become an 
isolated elliptical galaxy of moderate luminosity (highly unlikely), as we 
have not found any concentration of galaxies in its neighbourhood. 

\subsection{FS90\,110}

There is no radial velocity available for FS90\,110 and, as stated in 
Sect.\ref{K-corr}, it could be a background galaxy placed at $\sim$120 Mpc. 
However, it seems to be the most firm candidate for a cE galaxy. 

If it was an Antlia member, its projected 
separation from NGC\,3258 would be of the same order as those of M32 from M31, 
and NGC\,5486A from NGC\,5486. Moreover, it would be much closer to the 
dominant galaxy than NGC\,4486B to NGC\,4486, and its effective radius would 
be smaller than that of NCG\,5846A. 

It presents a high central surface brightness in the  
brightness profile obtained from the ACS image, and
its $(C-T_1)$ and $(V-I)$ colour profiles are quite flat, 
in agreement with what is found for M32. 
Its brightness profile shows two components in the equivalent radius 
range 2 -- 11.8 arcsec, that seems to be described by N $>$ 1 (i.e. n $<$ 1) 
S\'ersic indices. It is tempting 
to suggest that such N index for the outer component might 
be linked to a truncation of the profile \citep*{E08}, due to
a possible interaction with NGC\,3258. However, such hypothesis should be 
studied in the light of the confirmed membership status of FS90\,110.
Furthermore, it would be interesting to test if the outer component of 
confirmed cE galaxies displaying two components in their brightness profiles,
could be fitted by a S\'ersic law with N $>$ 1 (n $<$ 1). 

The variation of the ellipticity and position angle vs. radius displayed
by FS90\,110, resemble those found in M32. These changes occur at a  
galactocentric radius at which the outer component seems to begin to  
dominate, in agreement with what \citet{G02} found for M32.

FS90\,110 is the only FS90 cE candidate that follows the luminosity vs. mean 
effective surface brightness relation defined by bright ellipticals, which 
corresponds to the \citet{K77} scaling relation, towards fainter magnitudes, 
smaller radius, or higher mean effective surface brightness. It also 
seems to follow the Antlia members' colour-magnitude relation,
though it is located at the red border. 

The elongation and twisting of FS90\,110's outermost isophotes in the direction 
of NGC\,3258 are noticeable. That would be consistent with the 
extremely low surface brightness structure detected in the MOSAIC images, 
and confirmed with the FORS1 and ACS frames. Such kind of `bridge' that seems 
to link FS90\,110 with its bright partner, would be in agreement with the 
fact that most confirmed cE galaxies are located in the vicinity of brighter 
companions. In that sense, it is also remarkable the warped inner 
structure displayed in the ACS colour map, and 
confirmed through unsharp masks of FORS1. Quite similar stellar 
structures are found in numerical simulations \citep{B01,M06} as a consequence 
of galaxy interactions.

\subsection{Final remarks}

A galaxy that is interacting with a more massive partner 
will feel tidal forces most strongly  in its outskirts, 
while its central region will be less affected 
\citep[see, for example,][for a model of the interaction between M32 and 
M31]{B01}. As a consequence, it would not be surprising 
to observe asymmetric and `egg-shaped' outer isophotes, or/and detect 
low surface brightness stellar `tails', arising from the disruption of the 
outer regions of the satellite \citep{M06}. In particular, these faint 
structures could be observed either like one or two symmetric `tails',  
depending on projection effects \citep[see fig.\,4 in][]{M06}.
The distorted outermost isophotes of FS90\,110, and 
the extremely faint structure detected in the HST, VLT and CTIO images fit 
quite well in such a scenario.

The similarities in the ellipticity and position angle 
variations against radius displayed by FS90\,110 and FS90\,208, which arise 
at a similar galactocentric radius of $\sim 5.5$ arcsec (i.e. in their inner 
regions), are also remarkable. These kind of variations have already been 
detected in M32, although not so strongly. All these pieces of evidence lead us 
to speculate about the possibility that, an object similar to FS90\,208 might 
be the progenitor of a cE galaxy. Thus, it is tempting to look for possible 
links between these two kind of objects. 

We may think that a system with similar characteristics as FS90\,208 may 
lose its outermost regions due to the interaction with a 
bright companion, then becoming compact. As a consequence, it could also 
experience a redistribution of its stellar content, that might produce 
a warped inner structure, a higher central surface brightness, and an 
attenuation of its ellipticity and position angle variations with radius. 
Moreover, the bulge-to-disc ratio in such object could increase after losing 
its outer parts due to the interaction. 
With regard to this point, it should be noted that the inner component 
of FS90\,110 seems to be 3 times brighter than the outer one, while that of 
FS90\,208 is just 1.6 times brighter in relation with the outer component.

As an aditional point, we recall that the Antlia cluster seems to be 
particularly rich in S0 galaxies and FS90\,208 seems to be one of them.
If the evolutionary path of cE galaxies include this kind of objects, S0 rich 
clusters would arise as favorable environments for the formation of compact
galaxies. 

Dynamical simulations and the spectroscopically confirmed membership status
of FS90\,110 and FS90\,192, will help to test if any of the above statements 
is indeed plausible.  

\section*{Acknowledgements}
We would like to thank the referee, Igor Chilingarian, for his useful comments 
which helped to improve this paper. We are also grateful to S. A. Cellone for 
valuable discussions and for kindly reading the original manuscript.  
This research has made use of the NASA/IPAC Extragalactic 
Database (NED) which is operated by the Jet Propulsion Laboratory, California 
Institute of Technology, under contract with the National Aeronautics and 
Space Administration. 
This work was funded with grants from Consejo Nacional de Investigaciones 
Cient\'{\i}ficas y T\'ecnicas de la Rep\'ublica Argentina, Agencia Nacional de 
Promoci\'on Cient\'{\i}fica Tecnol\'ogica and Universidad Nacional de La Plata
(Argentina).
T.R. is grateful for support from the Chilean Center for Astrophysics, 
FONDAP No. 15010003.

\label{lastpage}

\begin{thebibliography}{}
%
\bibitem[\protect\citeauthoryear{Alonso et al.}{2003}]{A03} Alonso M. V., 
Bernardi M., Da Costa L. N., Wegner G., Willmer C. N. A., Pellegrini P. S.,
Maia M. A. G., 2003, AJ, 125, 2307 
%
\bibitem[\protect\citeauthoryear{Bassino, Richtler \& Dirsch}{Bassino et al.}{2008}]{B08} Bassino L. P., Richtler T., Dirsch B., 2008, MNRAS, 386, 1145 
%
\bibitem[\protect\citeauthoryear{Bekki et al.}{2001}]{B01} Bekki K., Couch W. J., 
Drinkwater M. J., Gregg M. D., 2001, ApJ, 557, L39
%
\bibitem[\protect\citeauthoryear{Binggeli, Sandage \& Tammann}{Binggeli et al.}{1985}]{B85} Binggeli B., Sandage A., Tammann G. A., 1985, AJ, 90, 1681
%
\bibitem[\protect\citeauthoryear{Binney \& Merrifield}{1998}]{BM98} Binney J., Merrifield M., 1998, in {\it Galactic Astronomy}, Princeton Series in Astrophysics, Princeton University Press, 184
%
\bibitem[\protect\citeauthoryear{Canterna}{1976}]{C76} Canterna R., 1976,
  AJ, 81, 228
%
\bibitem[\protect\citeauthoryear{Cellone \& Buzzoni}{2001}]{CB01} Cellone
  S. A., Buzzoni A., 2001, A\&A, 369, 742
%
\bibitem[\protect\citeauthoryear{Cellone \& Buzzoni}{2005}]{CB05} Cellone
  S. A., Buzzoni A., 2005, MNRAS, 356, 41
%
\bibitem[\protect\citeauthoryear{Chilingarian et al.}{2007}]{Ch07}
  Chilingarian I., Cayatte V., Chemin L., Durret F., Lagan\'a T. F., Adami
  C., Slezak E., 2007, A\&A, 466, L21
%
\bibitem[\protect\citeauthoryear{Chilingarian et al.}{2008}]{Ch08}
  Chilingarian I., Cayatte V., Durret F., Adami C., Balkowski C., Chemin L., 
  Lagan\'a T. F., Prugniel P., 2008, A\&A, 486, 85
%
\bibitem[\protect\citeauthoryear{Choi, Guhathakurta \& Johnston}{Choi et al.}{2002}]{choi02} Choi P. I., Guhathakurta P., Johnston K. V., 2002, AJ, 124, 310
%
\bibitem[\protect\citeauthoryear{Conselice, Gallagher \& Wyse}{Conselice et
    al.}{2002}]{Con02} Conselice C. J., Gallagher J. S., Wyse R. F. G., 2002,
  AJ, 123, 2246
%
\bibitem[\protect\citeauthoryear{Davidge}{1991}]{D91} Davidge T. J., 1991, AJ, 
102, 896
%
\bibitem[\protect\citeauthoryear{Djorgovski \& Davis}{1987}]{DD87} Djorgovski 
S., Davis M., 1987, ApJ, 313, 59
%
\bibitem[\protect\citeauthoryear{de Vaucouleurs et al.}{1991}]{dV91} de Vaucouleurs G., de Vaucouleurs A., Corwin H. G., Buta R. J., Paturel G., Fouque P., 
1991, Third Reference Catalogue of Bright Galaxies (RC3), Springer-Verlag, New
York
%
\bibitem[\protect\citeauthoryear{Dirsch, Richtler \& Bassino}{Dirsch et al.}{2003}]{dir03} Dirsch B., Richtler T., Bassino L. P., 2003, A\&A, 408, 929
%
\bibitem[\protect\citeauthoryear{Drinkwater et al.}{2001}]{D01}Drinkwater
  M. J., Gregg M. D., Holman B. A., Brown M. J. I., 2001, MNRAS, 326, 1076
%
\bibitem[\protect\citeauthoryear{Erwin, Pohlen \& Beckman}{Erwin et al.}{2008}]{E08}Erwin P., Pohlen M., Beckman J. E., 2008, AJ, 135, 20  
%
\bibitem[\protect\citeauthoryear{Ferguson \& Sandage}{1990}]{FS90}Ferguson
  H. C., Sandage A., 1990, AJ, 100, 1
%
\bibitem[\protect\citeauthoryear{Ferrarese et al.}{2006}]{F06}Ferrarese L.
  et al., 2006, AJSS, 164, 334
%
\bibitem[\protect\citeauthoryear{Forbes \& Forte}{2001}]{FF01}Forbes D. A., 
 Forte J. C., 2001, MNRAS, 322, 257
%
\bibitem[\protect\citeauthoryear{Fukugita, Shimasaku \& Ichikawa}{Fukugita et al.}{1995}]{F95}Fukugita M., Shimasaku K., Ichikawa T., 1995, PASP, 107, 945
%
\bibitem[\protect\citeauthoryear{Gavazzi et al.}{2005}]{G05} 
Gavazzi G., Donatti A., Cucciati O., Sabatini S., Boselli A., Davies J., Zibetti S., A\&A, 430, 411
%
\bibitem[\protect\citeauthoryear{Graham \& Worley}{2008}]{G08} 
Graham A. W., Worley C. C., 2008, MNRAS, accepted (arXiv:0805.3565)
%
\bibitem[\protect\citeauthoryear{Graham}{2002}]{G02} Graham A. W., 2002,
  AJ, 568, L13 (erratum 572, L121)
%
\bibitem[\protect\citeauthoryear{Geisler}{1996}]{gei96} Geisler D., 1996,
  AJ, 111, 480
%
\bibitem[\protect\citeauthoryear{Jedrzejewski}{1987}]{J87} Jedrzejewski 
R. I., 1987, MNRAS, 226, 747
%
\bibitem[\protect\citeauthoryear{Jerjen, Kalnajs \& Binggeli}{Jerjen et al.}{2000}]{J00} Jerjen H., Kalnajs A., Binggeli B., 2000, A\&A, 358, 845
%
\bibitem[\protect\citeauthoryear{Kormendy}{1977}]{K77} Kormendy J., 1977, ApJ, 218, 333
%
\bibitem[\protect\citeauthoryear{Lauer et al.}{1996}]{L96} Lauer T. R. et al.,
 1996, ApJ, 471, L79
%
\bibitem[\protect\citeauthoryear{Lauer et al.}{1998}]{L98} Lauer T. R., Faber
S. M., Ajhar E. A., Grillmair C. J., Scowen P. A., 1998, AJ, 116, 2263
%
\bibitem[\protect\citeauthoryear{Lisker et al.}{2006a}]{L06a} Lisker T., Glatt K., Westera P., Grebel E. K., 2006a, AJ, 132, 2432
%
\bibitem[\protect\citeauthoryear{Lisker, Grebel \& Binggeli}{Lisker et al.}{2006b}]{L06b} Lisker T., Grebel E. K., Binggeli B., 2006b, AJ, 132, 497
%
\bibitem[\protect\citeauthoryear{MacArthur, Courteau \& Holtzman}{MacArthur et al.}{2003}]{Mc03}MacArthur L. A., Courteau S., Holtzman J. A., 2003, ApJ, 582, 689
%
\bibitem[\protect\citeauthoryear{Mahdavi, Trentham \& Tully}{Mahdavi et al.}{2005}]{Ma05} Mahdavi A., Trentham N., Tully R. B., 2005, AJ, 130, 1502
%
\bibitem[\protect\citeauthoryear{Mateo}{1998}]{M98} Mateo M. L., 1998,
 ARA\&A, 36, 435
%
\bibitem[\protect\citeauthoryear{Mayer et al.}{2006}]{M06} Mayer L., 
Mastropietro C., Wadsley J., Stadel J., Moore B., 2006, MNRAS, 369, 1021
%
\bibitem[\protect\citeauthoryear{Mieske et al.}{2005}]{M05}Mieske S.,
  Infante L., Hilker M., Hertling G., Blakeslee J. P., Ben\'itez N., Ford H.,
  Zekser K., 2005, A\&A, 430, L25
%
\bibitem[\protect\citeauthoryear{Nakazawa et al.}{2000}]{nak00}
Nakazawa K., Makishima K., Fukazawa Y., Tamura T., 2000, PASJ, 52, 623
%
\bibitem[\protect\citeauthoryear{Nieto \& Prugniel}{1987}]{N87}Nieto J. -L.,
  Prugniel P., 1987, A\&A, 186, 30 
%
\bibitem[\protect\citeauthoryear{Pedersen, Yoshii \& Sommer-Larsen}{1997}]
{ped97}Pedersen K., Yoshii Y., Sommer-Larsen J., 1997, ApJ, 485, L17
%
\bibitem[\protect\citeauthoryear{Peletier}{1993}]{P93}
Peletier R. F., 1993, A\&A, 271, 51
%
\bibitem[\protect\citeauthoryear{Rose et al.}{2005}]{R05}
Rose J. A., Arimoto N., Caldwell N., Schiavon R., Vazdekis A., Yamada Y., 
2005, AJ, 129, 712
%
\bibitem[\protect\citeauthoryear{S\'anchez-Bl\'azquez, Gorgas \& Cardiel}{2006}]{SB06} S\'anchez-Bl\'azquez P., Gorgas J., Cardiel N., 2006, A\&A, 457, 823
%
\bibitem[\protect\citeauthoryear{Sandage \& Binggeli}{1984}]{SB84}
  Sandage A., Binggeli B., 1984, AJ, 89, 919
%
\bibitem[\protect\citeauthoryear{Sirianni et al.}{2005}]{S05}
  Sirianni M. et al., 2005, PASP, 117, 1049 
%
\bibitem[\protect\citeauthoryear{Schlegel, Finkbeiner \& Davis}{Schlegel et
    al.}{1998}]{S98}Schlegel D., Finkbeiner D., Davis M., 1998, ApJ, 500,
  525
%
\bibitem[\protect\citeauthoryear{S\'ersic}{1968}]{S68}
  S\'ersic J. L., 1968, Atlas de Galaxias Australes (C\'ordoba: 
  Obs. Astron., Univ. Nac. C\'ordoba)
%
\bibitem[\protect\citeauthoryear{Smith Castelli et al.}{2008}]{SC08}Smith 
Castelli A. V., Bassino L. P., Richtler T., Cellone S. A., Aruta C., Infante L., 
2008, MNRAS, 386, 2311 (Paper I)
%
\bibitem[\protect\citeauthoryear{Tonry et al.}{2001}]{ton01} Tonry J. L., 
Dressler A., Blakeslee J. P., Ajhar E. A.,Fletcher A. B., Luppino G. A.
Metzger M. R., Moore C. B., 2001, AJ, 546, 681
%
\bibitem[\protect\citeauthoryear{Trujillo et al.}{2001a}]{T01a} Trujillo I., 
Aguerri J., Cepa A. L., Guti\'errez C. M., 2001a, MNRAS, 321, 269
%
\bibitem[\protect\citeauthoryear{Trujillo et al.}{2001b}]{T01b} Trujillo I., 
Aguerri J., Cepa A. L., Guti\'errez C. M., 2001a, MNRAS, 328, 977
%
\bibitem[\protect\citeauthoryear{Ziegler \& Bender}{1998}]{ZB98}
  Ziegler B. L., Bender R., 1998, A\&A, 330, 819

\end{thebibliography}
\end{document}